%% file: main.tex
\documentclass{article}
\usepackage[letterpaper,top=2cm,bottom=2cm,left=3cm,right=3cm,marginparwidth=1.75cm]{geometry}
\usepackage{graphicx}
\usepackage{amsmath}
\usepackage{graphicx}
\PassOptionsToPackage{hyphens}{url}
\usepackage[colorlinks=true, allcolors=blue]{hyperref}
\usepackage{cleveref}
\usepackage{makecell}
\usepackage{booktabs}
\usepackage{tabularx}
\usepackage{ltablex}
\usepackage{caption}
\usepackage{subcaption}
\usepackage{stackengine}
\usepackage{multirow}
\usepackage{authblk}
\usepackage[table]{xcolor}
\usepackage{abstract}
\usepackage{breakcites}
\usepackage{amssymb}
\usepackage{graphicx}
\usepackage[T1]{fontenc}
\usepackage{palatino}

\title{When \textit{not} to target negative ties? \\ Studying competitive influence maximisation in signed networks}

\author[1,2,3]{Sukankana Chakraborty\footnote{corresponding author: \href{schakraborty@turing.ac.uk}{schakraborty@turing.ac.uk}}}
\author[3]{Markus Brede}
\author[3]{Sebastian Stein}
\author[4]{Ananthram Swami}

\affil[1]{Public Policy Programme, The Alan Turing Institute, British Library, UK.}
\affil[2]{Centre for Advanced Spatial Analysis, University College London, UK.}
\affil[3]{Electronics and Computer Science, University of Southampton, UK.}
\affil[4]{DEVCOM, Army Research Laboratory, USA.}

\date{}
\begin{document}

\maketitle


\begin{abstract}
We explore the influence maximisation problem in networks with negative ties. Where prior work has focused on unsigned networks, we investigate the need to consider negative ties in networks while trying to maximise spread in a population \textemdash particularly under competitive conditions. Given a signed network we optimise the strategies of a focal controller against competing influence in the network using two approaches \textemdash either the focal controller uses a sign-agnostic approach or they factor in the sign of the edges while optimising their strategies. We compare the difference in vote-shares (or the share of population) obtained by both these methods to determine the need to navigate negative ties in these settings.
More specifically, we study the impact of: (a) network topology, (b) resource conditions and (c) competitor strategies on the difference in vote shares obtained across both methodologies. We observe that gains are maximum when resources available to the focal controller are low and the competitor avoids negative edges in their strategy. Conversely, gains are insignificant irrespective of resource conditions when the competitor targets the network indiscriminately.
Finally, we study the problem in a game-theoretic setting, where we simultaneously optimise the strategies of both competitors. Interestingly we observe that, strategising with the knowledge of negative ties can occasionally also lead to loss in vote-shares.
\end{abstract}

\input{1a-introduction}
\input{1b-lit_review}
\input{2-methods}
\input{3-results}
\input{4-discussion}
\input{acknowledgments}
\input{A-appendix}

\bibliographystyle{apalike}
\bibliography{main}


\end{document}

%% file: 1a-introduction.tex
\section{Introduction}
\label{sec:introduction}
Opinions are predominantly influenced by individual beliefs, social interactions and information made available through external interventions \cite{myers1996social,watts2007influentials}. Of these, guiding collective opinions through external influence poses an interesting research problem which has far-reaching societal and commercial implications, such as trying to improve public health through controlled information campaigns \cite{wilder2018end}, campaigning in elections \cite{ranganath2016understanding} and marketing for product adoption \cite{watts2007viral} in populations.
The increasing reliance on social media and their pervasiveness in society today \cite{perrin2015social} has thus inspired a rich body of work dedicated to understanding how opinions evolve in social networks and how they can be optimally steered through external influence \cite{banerjee2020survey,noorazar2020recent}.
In particular, a number of papers explore this problem in social \cite{kiss2017mathematics,pastor2015epidemic}, political \cite{galam1999application} and economic settings \cite{easley2010networks,jackson2010social}. However, a majority of this literature strictly investigates friendship networks where influence propagates based on positive recommendations and endorsements \cite{newman2003structure}. Effects of negative relationships on opinion propagation have received limited attention and have been historically discounted from network dynamics, given their sparse presence \cite{offer2021negative} and association with avoidance behaviour \cite{harrigan2017avoidance}, i.e.~people who dislike (or distrust) each other are unlikely to communicate (or be connected within a network). However, the scope of anonymity and prevalence of fake profiles on social media platforms recently have made such ties increasingly ubiquitous \cite{bae2012sentiment,pfeffer2014understanding}, and even typical in many recommendation and trading networks \cite{guha2004propagation,leskovec2010signed}. Although this has initiated a lot of interest in understanding the impact of negative ties on network dynamics, there are several settings in this context that are yet to be explored \cite{offer2021negative}. 

In this paper, we consider negative ties as antagonistic relationships that negatively influence social neighbours, and persuade them to adopt an opposing position (or opinion).  
Such relationships pose a unique challenge when trying to maximise influence in a network. Relationships of distrust (negative reviews) on e-commerce platforms (e.g.~eBay) following below-par experiences, for instance, can negatively impact future transactions or communication \cite{borgs2010novel,chen2011influence}. Therefore, effective navigation of such ties is required when influencing a network externally. This is particularly the case in competitive environments, as neglectfully targeting individuals that propagate overall negative influence, can, in turn, facilitate the spread of undesirable opinions \cite{chen2018negative}.

In this paper, we demonstrate the need for a negative-tie aware approach while maximising influence in social networks under competitive settings. In doing so, we make the following contributions:
\begin{enumerate}
\item We modify voter dynamics for networks with negative ties, under competitive conditions and  subsequently present a negative-tie aware influence maximisation algorithm to optimally target networks in these settings.
\item We show that in a real-world network a controller can achieve an additional $9\%$ in vote-shares against a naïve competitor (with no knowledge of negative ties in the network).
\item We show how the effectiveness of a negative-tie aware approach varies with network topology, availability of budget and competitor strategy.
\item We provide analytical support for numerical methods and derive expressions for optimal allocations in large, arbitrary synthetic networks.
\item We present results for the game-theoretic setting where controllers actively optimise their strategies against one another. 
\end{enumerate}
The structure of the rest of the paper is as follows. In \cref{background} we discuss the related literature. We then modify the classical voter model in \cref{Model} to study opinion dynamics in signed networks. 
In \cref{bitcoin}, we investigate the problem in real-world settings, and we analyse it further using numerical methods in \cref{numerical}. In \cref{mean-field}, we present analytical support for our numerical results and further provide analytical expressions for optimal allocations in complex networks. Finally, in \cref{game-theory} we extend the problem to game-theoretic scenarios and we show how much a controller can gain from considering negative ties in influence maximisation exercises against unknown competitor strategies.

%% file: 1b-lit_review.tex
\section{Literature Review}
 \label{background}
Maximising influence or opinion shares in social networks through efficient allocation of external influence is a well-studied optimisation problem with wide applications \cite{easley2010networks,rogers2010diffusion}. Several approaches have been proposed to tackle this problem \cite{noorazar2020recent,li2018influence}. One of these is a computational method that maximises the number of converted (or infected) agents in a network, through cascading events triggered by an optimal set of initially infected individuals (or `seed' set) \cite{kempe2003maximizing}. Cascade models have been widely applied to maximise innovation adoptions in viral marketing \cite{domingos2001mining}, study the spread of infectious diseases \cite{cheng2020outbreak}, as well as rumour propagation \cite{tripathy2010study}. 

In a complementary dynamical approach, models from statistical physics \cite{castellano2009statistical} are often used to approximate emerging macroscopic behaviours in populations. In these settings, individuals repeatedly switch between different opinion states \cite{barrat2008dynamical} until the system reaches an equilibrium. Opinions expressed by agents are modeled as discrete \cite{clifford1973model,galam1999application,holley1975ergodic,krapivsky2003dynamics} or continuous \cite{deffuant2000mixing,hegselmann2002opinion} variables that change as they interact within the network. Of these, we employ the paradigmatic voter model \cite{clifford1973model,holley1975ergodic,sood2005voter} in our work. In the classical voter model, agents switch between binary states at a rate proportional to the fraction of opinion shares in their immediate neighbourhood. Our choice of model relies on its relevance \cite{PMID:28542409}, and the simplicity and tractability of its approach, that allows us to implement both analytical and numerical methods on complex networks \cite{redner2019reality}.

Dynamics in opinion models typically converge to an ordered consensus or fragmented states \cite{castellano2009statistical}. Since consensus is rarely achieved in the real world, it has motivated considerable efforts in understanding conditions that lead to opinion fragmentation in networks, with special attention to competitive settings \cite{hucompeting,prakash2012winner}. In such settings, controllers aim to exceed the competitor's share of opinions in the network at equilibrium.

Competitive opinion propagation has been extensively explored in the voter model using various forms of zealotry \cite{acemouglu2013opinion,mobilia2007role,yildiz2013binary}. Zealots are either biased individuals \cite{masuda2010heterogeneous,mobilia2003does}; radical agents impervious to influence from neighbours  \cite{barrat2008dynamical,kuhlman2013controlling,mobilia2007role,mobilia2015nonlinear} or external controllers unidirectionally targeting the network \cite{masuda2015opinion}. In this paper, we adopt a setting similar to that in \cite{masuda2015opinion}. However, in contrast to the norm of discretely targeting networks (where the most optimal seed set is identified), here we allow a continuous distribution of influence that targets nodes with varying intensities \cite{chakraborty2019competitive}. This modification widens the scope of application of this work to continuous resources such as money and time, while also significantly increasing the tractability of the optimisation problem.
 
 Much like zealots, contrarians too have been used to study competitive influence spread in the voter model \cite{gambaro2017influence,li2011strategy,masuda2013voter,zhong2005effects}. These are nonconformist agents opposing the majority view in their social neighbourhood. 
 We emulate this notion of disagreement in our work as a property of social relations, as opposed to that of agents, which is more realistic and a more general case of the contrarian model. 


  As largely reflected in society and real-world networks, social ties can often be negative, representing distrust and enmity \cite{leskovec2010signed}.
 So far, networks with signed edges have received limited attention within the vast body of work on opinion dynamics (see \cite{girdhar2016signed} for a detailed review). Most of this existing work explores the problem in threshold models \cite{hosseini2019assessing}. Principally, greedy heuristics are used in models of diffusion, such as the Independent Cascade (IC) \cite{ju2020new,li2014polarity,liu2019influence} and the Linear Threshold (LT) models \cite{he2019information,liang2019influence,shen2015influence}. 
 In addition, a slightly modified integrated page-rank algorithm for seed selection in signed networks was presented in \cite{chen2015influence}, while authors used simulated annealing for the same problem in \cite{li2017positive}. Work in \cite{srivastava2015social} discusses a comparative analytical approach to the same problem. Other generalised models of opinion propagation have also been used \cite{jendoubi2016maximizing}, including the voter model, where controllers seed the network optimally to maximise vote-shares at the steady-state \cite{li2013influence}.

 Our paper significantly extends the work in \cite{li2013influence} by framing the problem in more realistic, competitive and game-theoretic settings, which is different from the traditional single-controller influence maximisation problem studied in \cite{li2013influence}. We also further generalise the (discrete) optimisation problem presented in \cite{li2013influence} by allowing flexible distribution of influence in our model. Finally, we assume persistent influence from external controllers which diversifies the seeding process adopted in \cite{li2013influence}.

%% file: 2-methods.tex
\section{The opinion dynamics model}
\label{Model}
We consider a population of $N$ individuals interacting with each other in a social network through congenial (positive) or hostile (negative) relationships. We represent the structure of the network using a signed graph $G(V,E,W)$, where vertices $V = \{1,2,\ldots,N\}$ denote individuals in the population connected through a set of edges $E$ that depict social connections. Here $W \in R^{N \times N}$ denotes the corresponding signed weighted adjacency matrix, where any given element $w_{ij}$ illustrates the strength of influence an individual $i$ has on $j$.
The weight of an edge $w_{ij}$ determines the type of influence, positive or negative, that flows from $i \longrightarrow j$. A negative weight $w_{ij} < 0$ symbolises a negative edge and therefore implies negative influence from $i$ on $j$, whereas $w_{ij} > 0$ suggests node $j$ experiences positive influence from $i$. We consider directed networks, where edge weights $w_{ij} $ and $ w_{ji}$  are independent of each other, and it is possible that nodes experience different strengths and types (positive or negative) of influence from each other, i.e. $w_{ij} \neq w_{ji}$. 

We study the propagation of binary opinion states, A and B in the network, imposed by external controllers. 
Here we assume that controllers strictly exert positive influence on the network and external influence is expressed in terms of resource distribution vectors $\{p_{A},p_{B}\} \in \mathbb{R}^{N}_+$, where any element $p_{A,i} \geq 0$ (or $p_{B,i} \geq 0$) represents the strength of influence exerted by controller A (or B) on node $i$. The vectors are constrained linearly by the budget available to each controller, i.e. $\sum_{i}p_{A,i} = B_{A}$ (or $\sum_{i}p_{B,i} = B_{B}$). Unlike traditional models that assume a single-injection of influence at the start of the dynamics \cite{kempe2003maximizing,li2013influence}, here influence is applied continuously until the system reaches steady-state.

At any given point in time, individuals in the network strictly conform to either opinion states at a rate proportional to the strength of influence experienced by them. We assume opinion propagation follows voter dynamics, where at each time step a node $i$ chosen uniformly at random from the network updates its state by picking a source of influence (social neighbours or external controllers) to copy (or oppose in cases of negative strength). Node $i$ picks a neighbour $j$ with a probability $|w_{ji}|/(\textstyle\sum_{j \in {\cal N}_{i}^+ \cup {\cal N}_{i}^-} w_{ji} + p_{A,i} + p_{B,i})$\footnote{${\cal N}_{i}^+$ and ${\cal N}_{i}^-$ are the sets of positive and negative neighbours of node $i$.}, and either copies their state (when $w_{ji}>0$), or opposes them ($w_{ji}<0$). Similarly, node $i$ picks an external controller (say A) to copy, with the probability $p_{A,i}/(\textstyle\sum_{j \in {\cal N}_{i}^+ \cup {\cal N}_{i}^-} w_{ji} + p_{A,i} + p_{B,i})$. 


Assuming $x_{A,i}$ (or $x_{B,i} = 1 - x_{A,i}$) characterise the probability with which a node $i$ adopts opinion state A (or B), the rate at which $i$ transitions to opinion A is given by,
 
\begin{equation}
\frac{dx_{A,i}}{dt}=(1-x_{A,i}) \cdot \phi_{i}(A) -x_{A,i} \cdot \phi_{i}(B).
\label{Eq:1}
\end{equation}

The terms $\phi_{i}(A)$ and $\phi_{i}(B)$ indicate the fraction of total influence experienced by $i$ in favour of opinions A and B respectively and are given by,

\begin{align*}
\begin{array}{cc}
     & \phi_{i}(A) = \frac{\sum\limits_{j \in {\cal N}_i^+} w_{ji}x_{A,j}-\sum\limits_{j \in {\cal N}_i^-} w_{ji}(1-x_{A,j})+p_{A,i}}{\sum\limits_{j \in {\cal N}_i^+} w_{ji}-\sum\limits_{j \in {\cal N}_i^-} w_{ji}+p_{A,i}+p_{B,i}}; \\ \\
     & \phi_{i}(B) = \frac{\sum\limits_{j \in {\cal N}_i^+} w_{ji}(1-x_{A,j})-\sum\limits_{j \in{\cal N}_i^-} w_{ji}x_{A,j}+p_{B,i}}{\sum\limits_{j \in{\cal N}_i^+} w_{ji}-\sum\limits_{j \in {\cal N}_i^-} w_{ji}+p_{A,i}+p_{B,i}}.
\end{array}
\end{align*}
Here, A is our chosen focal controller. Similar expressions can be derived for controller B.

In signed networks, nodes experience both positive and negative influence from their neighbours. The collective positive influence is given by ${\displaystyle\sum}_{j \in {\cal N}_{i}^+} w_{ji}x_{A,j}$, and the total strength of negative influence (from neighbours in state B) is ${\displaystyle\sum}_{j \in {\cal N}_{i}^-} w_{ji}(1-x_{A,j})$.    
Edge weights $w_{ji}$ refer to incoming edges and allocations from external controllers A and B on node $i$ are $p_{A,i}$ and $p_{B,i}$ respectively. 

Given the above, the steady-state equation for the system can be evaluated by replacing $\frac{dx_{A,i}}{dt}=0$ in \cref{Eq:1} to obtain

\begin{equation}
   x_{A,i}^* = \frac{p_{A,i} - \sum\limits_{j \in {\cal N}_i^-}w_{ji} + \sum\limits_{j \in {\cal N}_{i}^+}w_{ji} x_{A,j} + \sum\limits_{j\in {\cal N}_{i}^-}w_{ji} x_{A,j}}{\sum\limits_{j\in {\cal N}_{i}^+}w_{ji} - \sum\limits_{j\in {\cal N}_{i}^-}w_{ji} + p_{A,i} + p_{B,i}}. 
    \label{steady-state}
\end{equation}

Here, $x_{A,i}^*$ is the probability a node $i$ has opinion state \emph{A} at equilibrium. 
For a network of size $N$ nodes, we obtain a system of $N$ equations, which can be given as follows, 

\begin{align*}
\quad
\begin{bmatrix}
L+diag(p_{A}+p_{B})
\end{bmatrix}
x_{A}^* &=
p_{A}-\vec{1}^T W^-,
\end{align*}
\begin{align}
\implies    X_{A} &= \frac{1}{N}\Vec{1}^{T} {x_{A}^*} = \frac{1}{N} \Vec{1}^{T}  [L+diag(p_{A}+p_{B})] ^{-1}(p_{A}-\vec{1}^T W^-).
    \label{optimisation}
\end{align}

$X_{A}$ in \cref{optimisation} denotes the total vote-share obtained by controller A at equilibrium. Assuming $W^+$ and $W^-$ are the weighted adjacency matrices of the positive and negative components of the network, the vector $\vec{1}^T W^-$ captures the total strength of negative influence on each node in the network. Here $L$ is an $N \times N$ matrix given by $L = diag(\vec{1}^T(W^+ - W^-)) - (W^+ + W^-)$. The diagonal elements represent the absolute sum of all edge weights of a node $i$, given by $L_{ii} = {\displaystyle\sum}_{j \in {\cal N}_{i}^+} w_{ji} - {\displaystyle\sum}_{j \in {\cal N}_{i}^-} w_{ji}$ and off-diagonal elements are $L_{ij} = -w_{ji}$. For unweighted graphs $L = diag(D) - (A^+ - A^-)$, where $D$ is the degree-vector\footnote{diag(D) is an $N \times N$ matrix where the diagonal elements capture the degrees of nodes in the network.}. $A^+$ and $A^-$ are the respective adjacency matrices of the positive and negative components. Note that since $[L + diag(p_{A} +p_{B})]$ is diagonally-dominant, it is invertible and we can therefore use \cref{optimisation} to determine solutions for $X_{A}$. 

The formal optimisation problem can then be stated as

\begin{align}
    p_{A}^* = \text{argmax}_{p_{A} \in \mathcal{P}} \quad X_{A}^*(L,p_{B}),
    \label{formal_opti}
\end{align}
where $\mathcal{P}$ is a set of all possible allocations $p_{A}$ such that $0 \leq p_{A,i} \leq B_{A}$ ($\forall i \in \{1,2,\ldots,N\}$) and $\sum\limits_{i=1}^N p_{A,i} = B_{A}$.

For a passive opponent B (where $p_{B}$ is fixed and known), controller A maximises their opinion shares using Eq. \ref{formal_opti}. 
Closed-form solutions can be readily obtained in networks with simplified structures (e.g. star networks) by solving the set of equations outlined in \cref{optimisation}. To do this, we first determine the gradient $\nabla_{p_{A}} X_{A}$ by differentiating \cref{optimisation} wrt to the allocation vector $p_{A}$ as

\begin{align}
    \nabla_{p_{A}} X_{A} = \frac{1}{N} \Vec{1}^{T} [L+diag(p_{A}+p_{B})] ^{-1} (I - diag(x_{A})),
    \label{gradient}
\end{align}

and then solve $\nabla_{p_{A}} X_{A}=0$ to obtain the optimal allocation $p_{A}^*$ that yields maximum vote-shares $X_{A}^*$. However, obtaining analytical solutions for $\nabla_{p_{A}} X_{A}=0$ in larger, more complex networks can be considerably challenging. In which case we use local search techniques such as gradient ascent (as they have been shown to work well in similar settings \cite{lynn2016maximizing,romero2021shadowing}) to determine optimal allocations in arbitrary networks. More specifically, we employ the gradient algorithm proposed in \cite{romero2021shadowing}, by first modifying it to fit our purpose. Note that the problem setting considered in \cite{romero2021shadowing} resembles our research problem closely, as both studies explore the competitive influence maximisation problem in the voter model under continuous allocations, with the exception that the model given in \cite{romero2021shadowing} only applies to unsigned (or positive) networks. This yields a slightly different expression for vote-shares and does not contain the term $-\vec{1}^T W^-$ (as in \cref{optimisation}). Despite this difference, the expression for gradient is the same as \cref{gradient}, given that $-\vec{1}^T W^-$ is independent of controller allocations $p_{A}$. Thus we replace the expression for vote-shares in the algorithm (from \cite{romero2021shadowing}) with \cref{optimisation} and then use it to obtain optimal allocations. We label this approach $GA$.

In order to determine the effectiveness of the negative-tie aware method, we compare it against the algorithm that considers all edges to be positive, therefore discounting any negative influences in the network dynamics, when solving the influence maximisation problem. We then measure the impact of taking a negative-tie aware approach on vote-shares and quantify the gain or loss. There are two broad cases where controllers might be negative-tie agnostic: (i) where they cannot observe negative ties (in cases of under-representation) or (ii) where mistake them for positive edges (misrepresentation) \cite{li2013influence}. For the sake of simplicity, we avoid considering instances where controllers can partially observe negative edges in the network.

We are keen to determine instances where a controller would gain the most from adopting a negative-tie aware approach and we find that misrepresenting negative-ties consistently performs worse as a strategy in comparison to under-representation of negative-ties (see \cref{appendix-rem} for more details). Thus in the rest of the paper, we compare our proposed method to the approach where negative ties are misrepresented (i.e. all edges are assumed to be positive)\footnote{In undirected networks, both approaches \textemdash misrepresentation and under-representation of negative ties \textemdash would yield equal vote-shares.}.

To determine an expression for vote-shares in networks where all edges are positive, we modify \cref{steady-state} that yields an expression for vote-shares, given by $X_{A}^{(+)}$ (identical to the expression for vote-shares in \cite{romero2021shadowing}). We then directly apply the algorithm in \cite{romero2021shadowing} to obtain optimal allocations $p_{A}^*$. For ease of notation, we label this approach $GA^{(+)}$.





%% file: 3-results.tex
\section{Relative gain in opinion shares in a real-world network}
\label{bitcoin}
As a first step we explore the extent to which knowledge of negative ties benefits a controller in a real-world influence maximisation setting. To do this, we consider a real-world Bitcoin trust network \cite{kumar2016edge,kumar2018rev2}, where edges are labelled with positive and negative signs. The network is formed by members of the Bitcoin OTC platform who rate each other based on their interactions and transaction experiences. Ratings (or votes) show the extent to which a member is trusted in the community, and hence it forms their reputation on the platform for future transactions. 
Here we examine the extent to which relationships of distrust, if unobserved, can jeopardise influence maximisation efforts.   

The Bitcoin network is a weighted-directed graph of 5.8K nodes. Edges are weighted between -10 and +10, where -10 implies ultimate distrust and +10 stands for complete trust. 
For our experiment, we focus on the single largest connected component of size 4.7K nodes that contains $\approx 94\%$ of all edges. 
Within this component, $\approx 8.6\%$ of all edges are negatively weighted. In our model, influence flows in the direction opposite to that of rating, i.e. if a user $i$ gives a score of +10 to user $j$, we assume that user $i$ would copy the opinion state of user $j$ with strength $w_{ji}=10$. In case of a negative score (-10), user $i$ adopts the opposing state with strength $w_{ji}=-10$. 

We then run $GA$ and $GA^{(+)}$ on the network to measure the advantage of using a negative-tie aware optimisation process. This is obtained as the relative gain in vote-shares $[X_{A}^{(GA)}/X_{A}^{(GA+)} - 1]$ and yields the the fraction of vote-shares a controller can gain from a negative-tie aware approach. In the rest of the paper, we use this metric as a standard to measure the advantage of navigating negative ties in these settings.

We assume that the competing controller is naïve and targets the network indiscriminately with uniform allocations $p_{B,i} = 1$ (budget $B_{B} = N$). Intuitively, it would make sense to assume that the $GA$ algorithm would perform better in comparison to $GA^{(+)}$ if controller B was aware of negative edges and adopted a negative-tie aware strategy. By assigning an indiscriminate strategy to B we control for any effect of the competitor strategy on vote-share gains and tease out the gain in vote-shares that is obtained from following a negative-tie aware approach. Later in the paper, we explore the effect of competitor strategy on vote-share gains.

\begin{figure}
  \begin{subfigure}[b]{0.475\textwidth}
  \centering
    \includegraphics[width=\textwidth]{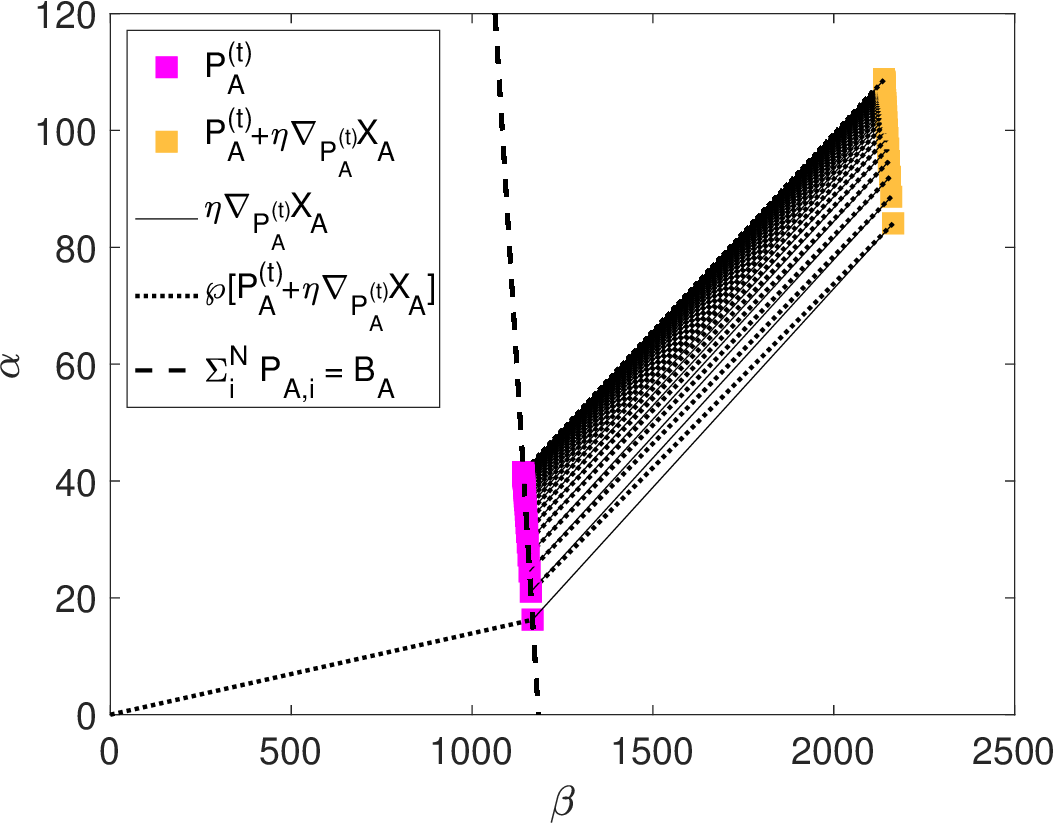}
    \caption{}
    \label{Fig-1a}
  \end{subfigure}
   \hspace{1em}
  \begin{subfigure}[b]{0.475\textwidth}
  \centering
    \includegraphics[width=\textwidth]{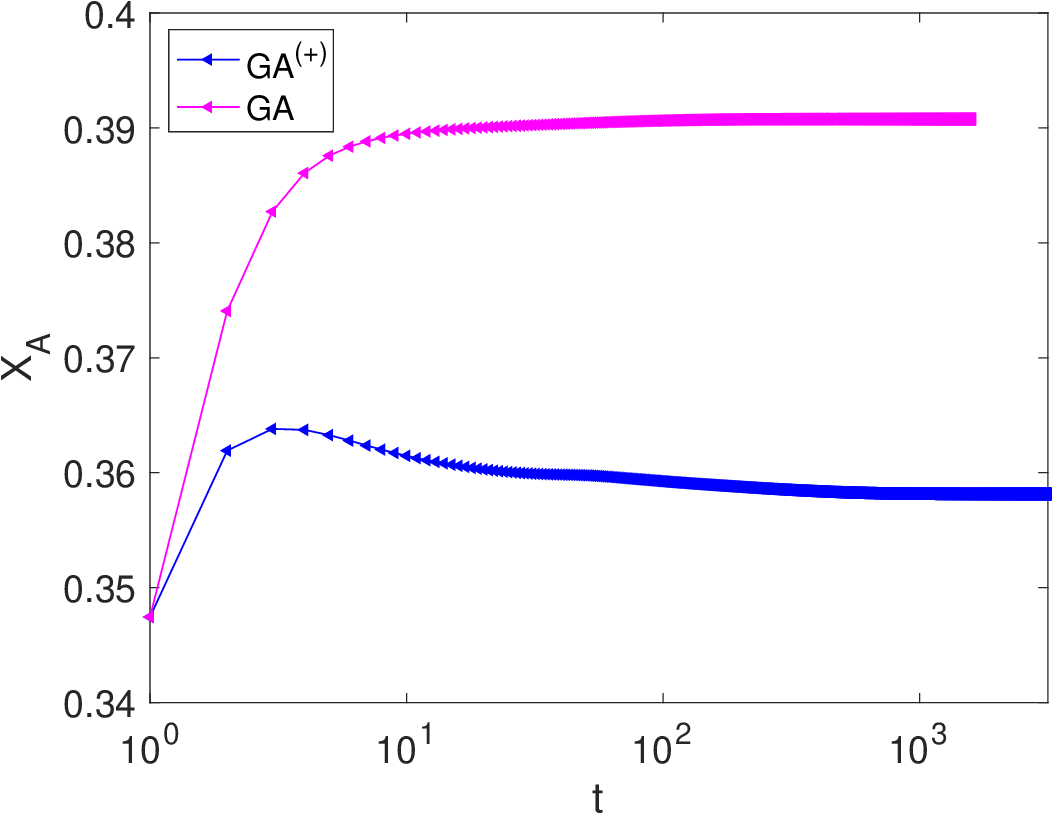}
    \caption{}
    \label{Fig-1b}
  \end{subfigure}
  \caption{Simulations are performed on the single-largest connected component ($N = 4734$ nodes) of the Bitcoin OTC network. Panel (a) shows step-wise allocations to nodes in the $GA$ algorithm. The axes represent the total allocations $\alpha$ to nodes with negative links and the total allocations $\beta$ to nodes with strictly positive edges.   Panel (b) shows the change in vote-shares ($X_{A}$) 
  obtained by both algorithms ($GA$ and $GA^{(+)}$) as a function of the number of iterations performed. Controller B passively targets all nodes in the network, uniformly, with a budget $B_{B} = N$. Controller A here has only a quarter of the resources available to B, $B_{A} = B_{B}/4 = N/4$. Algorithms are terminated using an approximation factor of $\mu = 10^{-7}$. The learning rate is initialised at $\eta=N$.}
  \label{Fig-1}
\end{figure}

For our simulations, we initialise $p_{A}$ with a uniform random vector (constrained by $B_{A}$) and use the gradient to improve the distribution vector until it converges to an optimum. The learning rate is set as $\eta=N$ and 
an approximation factor of $\mu = 10^{-7}$ is used to terminate the algorithms. 
Simulation results are summarised in \cref{Fig-1} for the instance $B_{A} = N/4$ as the budget ratio $B_{A}/B_{B}=0.25$ yields the highest gain in vote-shares (see \cref{appendix-rem}).

In \cref{Fig-1a} we illustrate the performance of the $GA$ algorithm on the Bitcoin OTC network. In particular, we show the convergence of the algorithm over the final $500$ steps. Note how the allocation vector $p_{A}^{(t)}$ is incrementally changed at each time step $t$ by stepping $\eta\nabla_{p_{A}^{(t)}}X_{A}$ in the direction of the gradient, where $\eta$ is the step-length. The resulting vector is then projected back onto an $N$-simplex to preserve the budget constraint $\sum_{i}^{N} p_{A,i} = B_{A}$.  

In \cref{Fig-1b} we illustrate the evolution of vote-shares using both algorithms over time $t$. We find that when using $GA$ the vote-share monotonically increases to a final configuration of $X_{A}^{*(GA)} \approx 0.3908$. While $GA^{(+)}$ has a suboptimal outcome and converges to $X_{A}^{*(GA+)} = 0.3582$, resulting in an overall gain of $\approx 9\%$ for $GA$. Thus highlighting that even in the presence of a naïve competitor, it is useful to consider negative edges when maximising influence in a signed network.

\section{Numerical analysis}
\label{numerical}
We now analyse the problem more thoroughly under controlled settings.
In the model introduced above, there are three salient factors that can affect the difference in performance of both algorithms (\emph{GA} and ${GA^{(+)}}$). These are the (a) topology of the network ($L$ and $L^{(+)}$), (b) resource conditions ($B_{A}/B_{B}$) and (c) competitor strategies ($p_{B}$). Below, we systematically explore the effect of each of these factors on the gain in vote-shares.  

\subsection{Role of network topology}
\label{hetero}

First, we investigate the effect of network topology. Specifically, we examine how the placement and distribution of negative edges in the network can affect gain in vote-shares. To do this we generate synthetic networks that vary in terms of the distribution of negative ties. For our analysis, we consider undirected networks of size N=1000 nodes with an average positive degree ${\textstyle\sum}_{i}^N k_{a,i}/N = \langle k_{a} \rangle =16$ and average negative degree ${\textstyle\sum}_{i}^N k_{b,i}/N =\langle k_{b} \rangle =4$. Here $k_{a,i} = {\textstyle \sum}_{i}^{{\cal N}_{i}^+} w_{ji}$ and $k_{b,i} = - {\textstyle \sum}_{i}^{{\cal N}_{i}^-} w_{ji}$ represent the total positive strength and the total negative strength on a node $i$. We set $\langle k_{a} \rangle$ and $\langle k_{b} \rangle$ such that only $20\%$ of all edges in the network are negative. We keep this fraction low, first and foremost, to imitate real-world networks \cite{leskovec2010signed}, and secondly to determine the importance of considering negative ties even when they are sparingly present. 

We generate synthetic signed networks for our analysis by first generating a network of N=1000 nodes with all positive edges where $\langle k_{a} \rangle =16$. We then add negative edges to this network to form our signed network. As we are interested to determine how the structure of the networks impact the efficiency of the negative-tie aware approach, we vary the distribution of negative edges in the network by using a parameter $p$ which determines the fraction of nodes in the network that have negative edges. Thus for lower values of $p$, negative edges are concentrated on a few nodes, whereas for $p=1$, every node in the network has at least one negative edge.

Signed networks for our experiment are synthesised by merging independently constructed positive and negative graphs. Merging the two graphs helps us control the distribution and placement of the negative edges.
The positive component has a size of N=1000 nodes with $\langle k_{a} \rangle =16$. The size of the negative graph is given by $p \cdot N$, where $p$ indicates the fraction of nodes in the network with a negative edge. Once the negative component is generated it is superimposed on the positive graph, by first choosing $p\cdot N$ nodes from the positive graph and then aligning them with the nodes of the negative graph to combine both subgraphs, i.e. each node from the negative graph is mapped to a node in the positive component, such that the node retains both positive and negative edges (from both subgraphs). The placement of negative edges is controlled by the way in which nodes are chosen while merging the networks\footnote{Discussed later in the section.}. Note that while combining the subgraphs, if two nodes have both a positive and a negative edge between them, we then keep the positive edge and ignore the negative tie. 

To perform our analysis on a broad spectrum of network structures, we vary the heterogeneity of degree distribution in the subgraphs. In this instance, subgraphs can be either homogeneous regular graphs \cite{newman2011structure} or heterogeneous core-periphery type networks \cite{rombach2017core} with bi-modal degree distributions\footnote{We use the configuration model \cite{newman2018networks} to generate the network structures.}. We then combine positive and negative subgraphs with different degree-distributions to generate three distinct classes of networks for our experiment.

\textbf{Reg-Reg :} Our benchmark is a homogeneous network where both positive and negative subgraphs are regular graphs (Reg-Reg). The positive component is a $k^+$-regular graph where $k^+=\langle k_{a} \rangle = 16$. For the negative subgraph, we generate homogeneous networks of size $p \cdot N$ and degree $k^- = \langle k_b \rangle / p$, where $p$ is varied between $0.1 \leq p \leq 1$ \footnote{Where $k^{-} \notin \mathbb{Z_+}$, we generate a $k^{-}$-regular graph and add ties to a fraction of randomly selected nodes to increase their negative degree by 1 ($k^{-} + 1$). This roughly preserves the average negative degree and consequently the percentage of negative edges in the network.}. Subgraphs are then combined by merging nodes from the negative subgraph with randomly chosen nodes from the positive subgraph.

\textbf{CP-Reg :} In this class of networks, the positive graphs have a heterogeneous core-periphery structure where half of all nodes (N/2) have high-degrees $k_{a,high} = 30$ and the rest have low-degrees $k_{a,low} = 2$. The negative component here has a regular or pseudo-regular structure (as described in the earlier section). 
We adopt three different approaches to merge the positive and the negative graph components, which varies the placement of negative ties in the network. 
Negative edges are either preferentially added to: (i) high-degree nodes in the positive component (i.e. only nodes with high-degrees in the positive component are chosen for merging (labelled CP-Reg-High)), (ii) low degree nodes in the positive component (CP-Reg-Low) or (iii) negative edges are randomly placed (CP-Reg-Rand).

\textbf{Reg-CP :} Our final set of networks have a homogeneous positive component and is a $k^{+}$-regular graph where $k^{+} = \langle k_{a} \rangle = 16$, and the negative component has a core-periphery structure with $\langle k_{b} \rangle = 4$. Nodes in the negative subgraph are split equally into two classes: (i) high-degree nodes ($k_{b,high} = 2(k_{b}/p-1)$) and (ii) low-degree nodes ($k_{b,low} = 2$). Here $p$ ranges from $0.15 \leq p \leq 1$ as it is not feasible to form core-periphery structures for $p=0.1$.

\begin{figure}
  \begin{subfigure}[b]{0.475\textwidth}
  \centering
    \includegraphics[width=\textwidth]{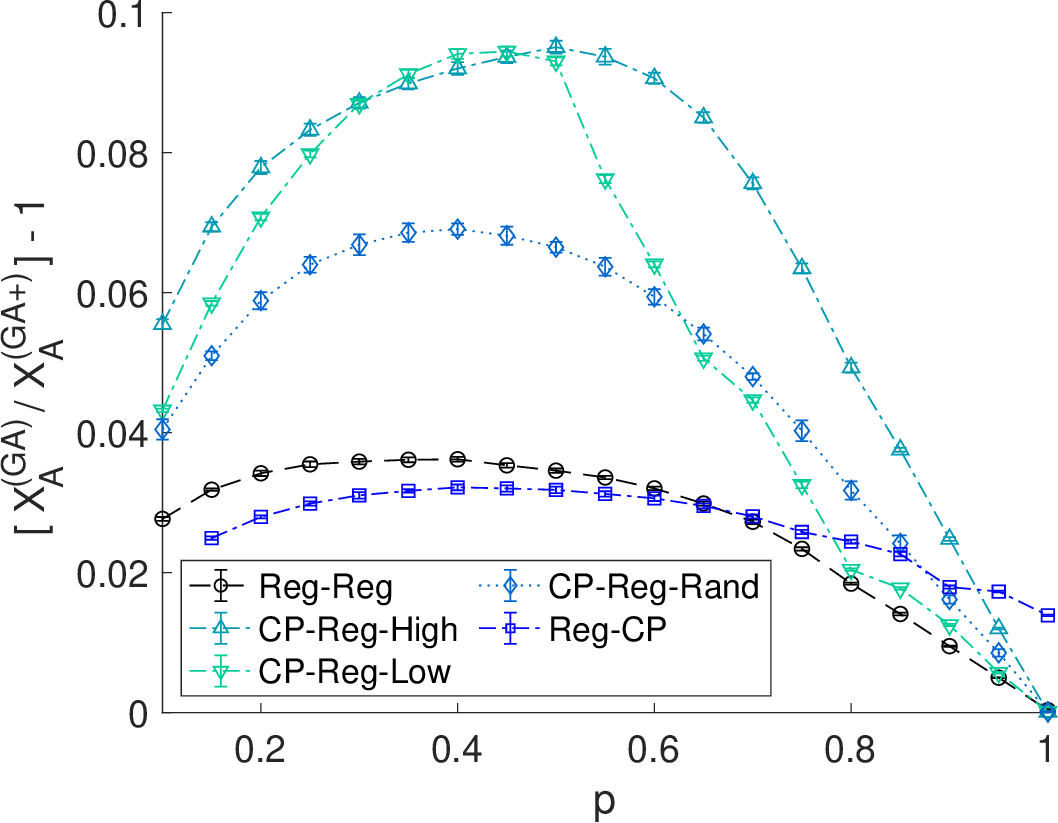}
    \caption{}
    \label{Fig-2a}
  \end{subfigure}
   \hspace{1em}
  \begin{subfigure}[b]{0.475\textwidth}
  \centering
    \includegraphics[width=\textwidth]{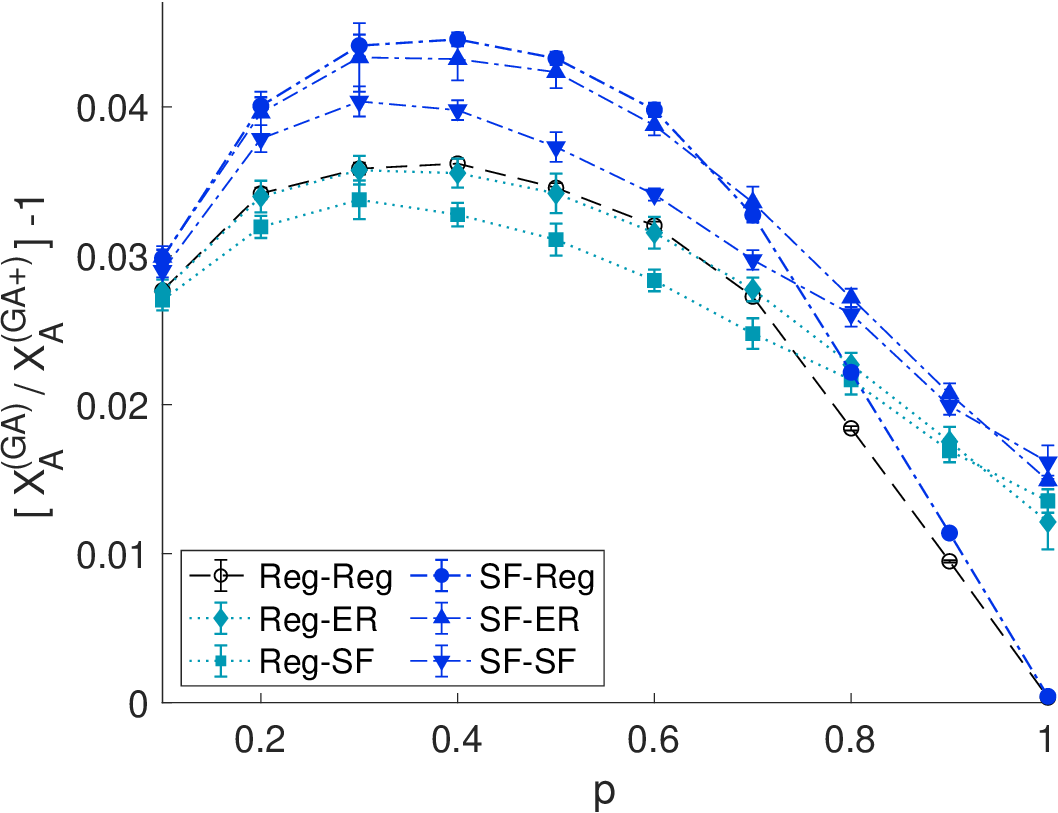}
    \caption{}
    \label{Fig-2b}
  \end{subfigure}
  \caption{Figure showing relative gain in vote-shares as the distribution of negative edges (p) is varied. Panel (a) shows the effect of heterogeneity of positive (CP-Reg) and negative edges (Reg-CP), compared against networks where both positive and negative components are regular (Reg-Reg). Similarly, panel (b) examines the role of heterogeneity in other commonly studied networks. The heterogeneity of negative edges is varied as the negative graph is changed from a regular (Reg-Reg) to a random (Reg-ER), and finally, a scale-free (Reg-SF) network. Heterogeneity of the positive component is achieved by replacing the regular positive graph (Reg-Reg) with a scale-free network (SF-Reg). 
  For all simulations, networks of size $N=1000$ with $\langle k_{a} \rangle = 16$ and $\langle k_{b} \rangle = 4$ are used. Results are averaged over 10 instances and error bars depict 95\% confidence intervals. Note that the panels are scaled differently for clarity.}
  \label{Fig-2}
\end{figure}

For each class of networks discussed above, we run simulations over 10 networks and summarise the results in \cref{Fig-2a}. We assume controller B is unaware of negative edges and passively targets each node with unit resource $p_{B,i}  = 1$ $\forall i \in \{1,2,..,N\}$ and controller A has a budget $B_{A} = N$.

\textbf{Scale-free and random networks}
We also extend our study to commonly studied Erdös-Rényi (ER) graphs \cite{erdHos1960evolution} and Barabási-Albert (BA) networks \cite{barabasi2013network}. Here the positive subgraph is either a regular graph (Reg) or a scale-free network (SF), and the negative component varies between a regular graph (Reg), a random graph (ER) and a scale-free network (SF).
We assume heterogeneity is analogous to degree variance \cite{bell1992note}, and is gradually increased in the negative components by replacing regular networks (Reg-Reg) with random graphs (Reg-ER) and then with scale-free networks (Reg-SF). In each case, the placement of negative edges is random. Finally, as before, results are compared against the homogeneous Reg-Reg graphs where both positive and negative components are k-regular networks, i.e. $ k = k^{+} + k^{-}=\langle k_{a} \rangle + (\langle k_{b} \rangle/p)$. 
We present the results in \cref{Fig-2b}.  

\cref{Fig-2} highlights the effect of degree heterogeneity on the gain in vote-shares as the distribution of negative edges $p$ is varied.
In general, we find that the relative gain in vote-shares first increases and then decreases as negative edges are distributed more evenly in the network. For networks with a homogeneous negative component the relative gain reduces to $0$. An analytical explanation of this effect is provided in \cref{gain0}.

In \cref{Fig-2a}, we find that networks with a positive core-periphery component and a regular negative subgraph (CP-Reg) yield the highest relative gain in vote-shares. Furthermore, relative gain is higher when placement of negative ties are degree-correlated (as in CP-Reg-High and CP-Reg-Low) compared to when randomly placed in the network (CP-Reg-Rand). 
A controller achieves maximum relative gain in vote-shares when negative edges are confined to nodes of a single degree-type ($p=0.5$), high-degree (CP-Reg-High) or low-degree nodes (CP-Reg-Low).
The maximum gain in vote-shares is $9.5\%$ when only hub nodes (with high positive degrees) have negative ties (CP-Reg-High). Similarly gain is $9.4\%$ when only peripheral nodes (with low positive degrees) have negative ties (CP-Reg-Low) . 
We observe similar trends in \cref{Fig-2b} where a maximum gain of $4.5\%$ is obtained in networks with a heterogeneous scale-free positive component and a homogeneous regular graph (SF-Reg). This result is close to the $6.9\%$ gain observed in CP-Reg-Rand networks where negative edges are randomly added to nodes (as in SF-Reg). We also find that substituting negative regular subgraphs with more heterogeneous components, such as a random graph (ER) or a scale-free network (SF), progressively decreases gain in vote-shares. Therefore suggesting that gains are high only in networks where nodes vary significantly based on their positive degrees and where negative ties are uniformly spread to about $50\%$ of the population.



\subsection{Role of resource conditions and competitor allocations}
\label{resources}
We now examine how resource availability ($B_{A}/B_{B}$) impacts the relative gain in vote-shares. 
Preliminary results indicate insignificant gains ($\leq1\%$) when the controller has more budget than its competitor $B_{A,i} > B_{B,i}$. This shows that controllers with large budgets can only marginally gain from exploiting their knowledge of negative ties in the network and therefore we focus our analysis on conditions of lower resources, i.e. $B_{A,i} \leq B_{B,i}$, in the remainder of the section.

To determine if resource conditions can further exacerbate the difference in vote-shares for $GA$ and $GA^{(+)}$ and to make our analysis more effective, we restrict our investigation to topologies that yielded the highest gains in the previous section. Thus, we choose networks with heterogeneous positive components and homogeneous negative subgraphs (CP-Reg). Simulations are performed only on CP-Reg-High networks, for conciseness.  

The budget for controller B is fixed at $B_{B,i} = 1$ per node for the rest of the experiments in this section. Controller A has a per node budget of $0.05 \leq B_{A,i} \leq B_{B,i}$.
For simplicity, controller B continues to target the network passively. We consider three strategies for controller B to allocate resources: (i) avoid nodes with negative edges, (ii) only target nodes with negative edges, and (iii) target all nodes uniformly. Note that in each case, B evenly distributes its resources over all targeted nodes. In the first two cases, we assume controller B is aware of negative edges, and decides to either target or avoid them. This allows us to examine how the gain in vote-shares depends on the adversary's knowledge of the network structure.

\begin{figure}
  \begin{subfigure}[b]{0.45\textwidth}
    \includegraphics[width=\textwidth]{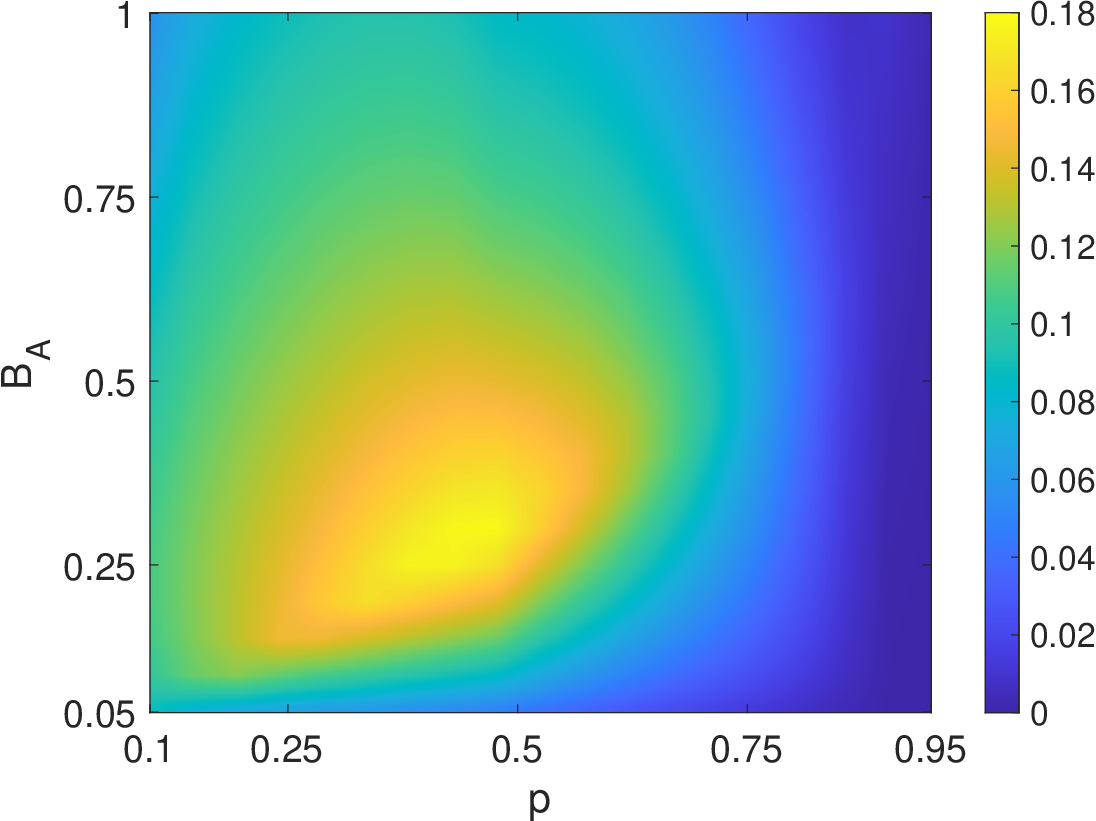}
    \caption{}
    \label{Heatmap-1}
  \end{subfigure}
\hspace{2em}
  \begin{subfigure}[b]{0.45\textwidth}
    \includegraphics[width=\textwidth]{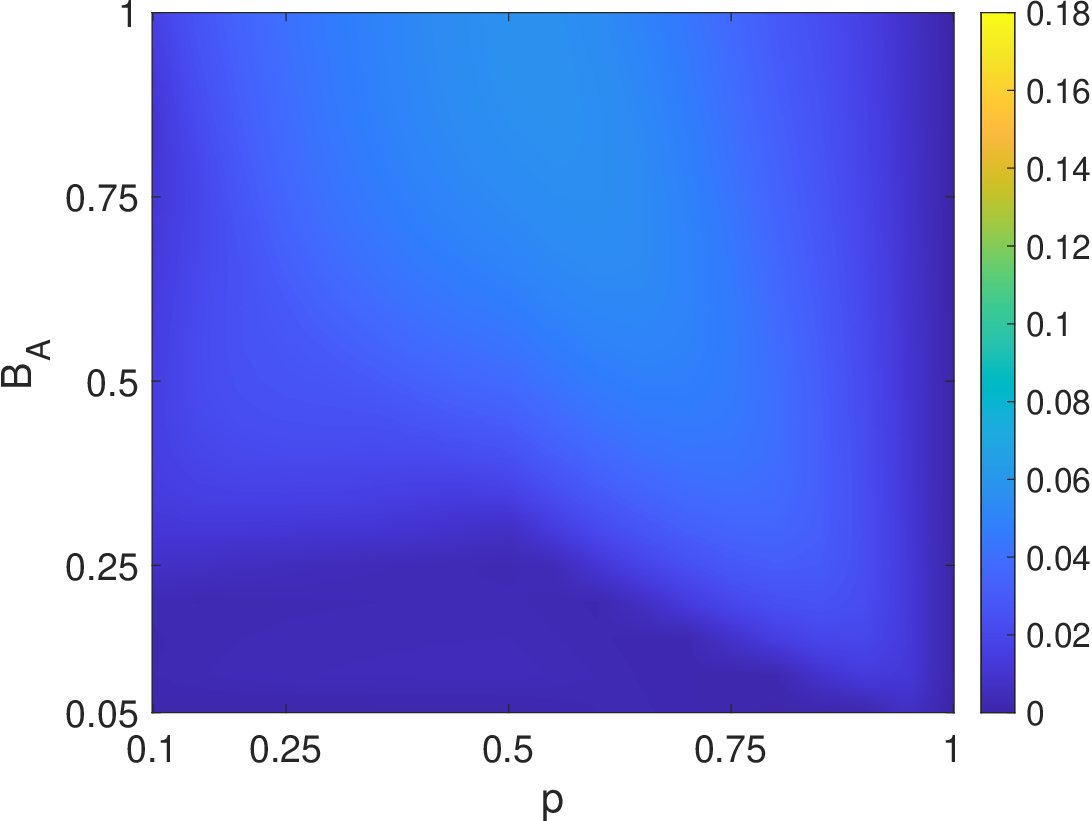}
    \caption{}
    \label{Heatmap-2}
  \end{subfigure}

  \begin{subfigure}[b]{0.45\textwidth}
    \includegraphics[width=\textwidth]{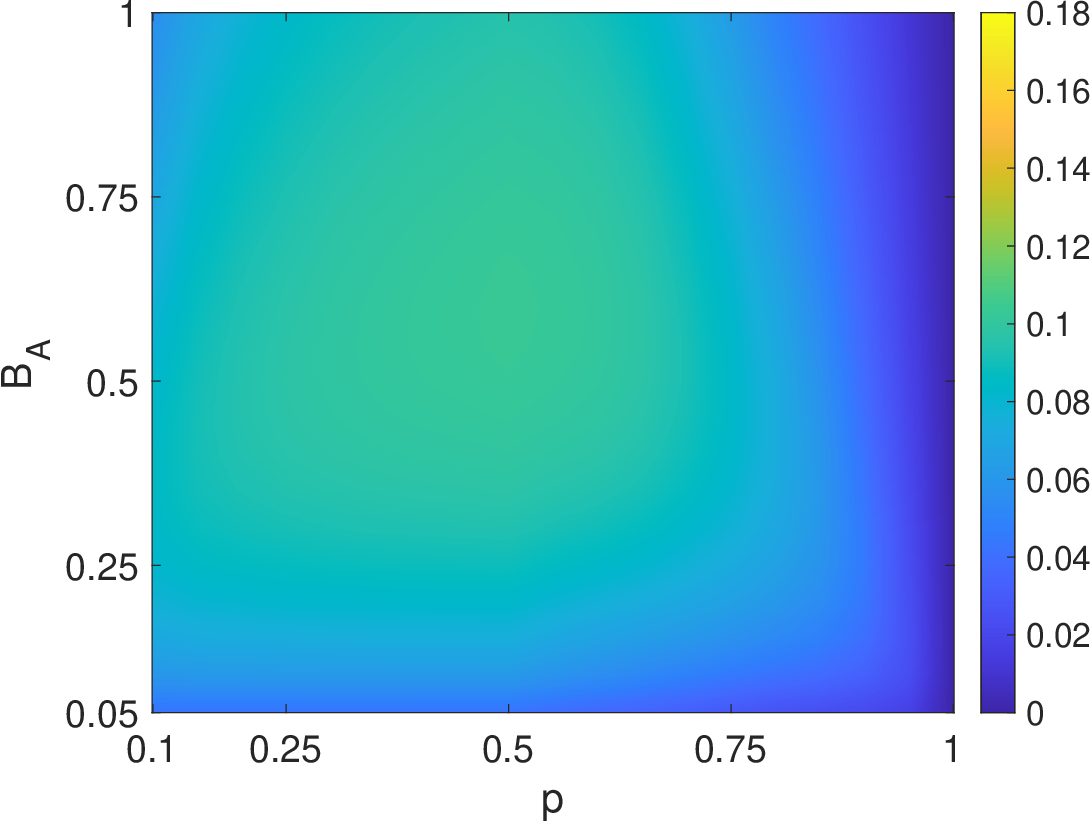}
    \caption{}
    \label{Heatmap-3}
  \end{subfigure}
   \hspace{2em}
  \begin{subfigure}[b]{0.45\textwidth}
    \includegraphics[width=\textwidth]{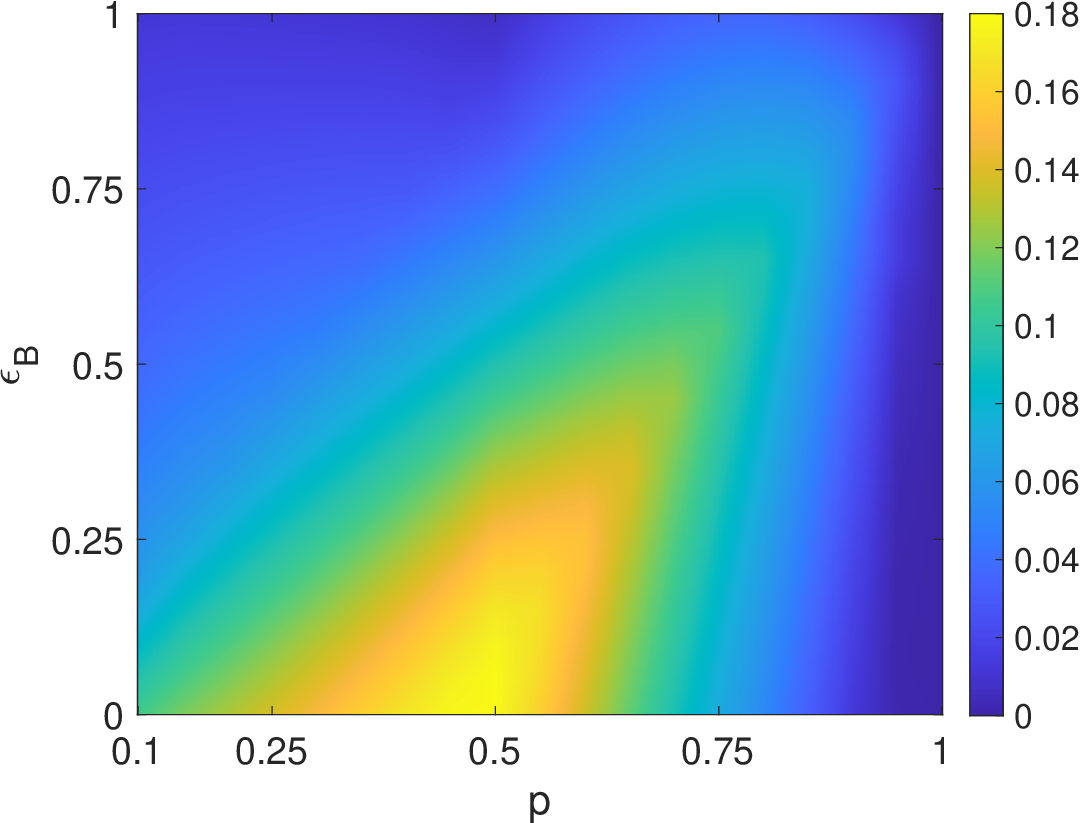}
    \caption{}
    \label{Heatmap-4}
  \end{subfigure}
\caption{Panel showing relative gain in vote-share as $p$ and budget ratios are varied against different competitor strategies. We examine three cases, controller B: (a) avoids nodes with negative ties, (b) targets nodes with negative ties and (c) targets all nodes uniformly. For all cases, controller B has a fixed budget $B_{B}=1$.
Lastly, in (d) we quantify gain in vote-share as controller B changes its strategy by varying the fraction of budget $\epsilon_{b}$ used to target nodes with negative edges. Here both controllers have fixed budgets where $B_{A} = 0.3$ and $B_{B}=1$. 
Results are averaged over 10 CP-Reg-High networks of size $N = 1000$ nodes and $\langle k_{a} \rangle =16$, $\langle k_{b} \rangle = 4$. }
\end{figure}

Gain in vote-shares for each case is presented in \cref{Heatmap-1,Heatmap-2,Heatmap-3}. Here \cref{Heatmap-1} illustrates the gain in vote-shares when controller B avoids nodes with negative edges. In \cref{Heatmap-2}, controller B exclusively targets nodes with negative ties and in \cref{Heatmap-3} controller B targets the network uniformly.

From the numerical experiments presented in \cref{Heatmap-1,Heatmap-2,Heatmap-3}, we make the following observations. First, we find that controller A can lose considerable vote-shares by not adopting a negative-tie aware approach when controller B avoids negative edges.
On the other hand, the difference in vote-shares is the least when controller B deliberately targets nodes with negative edges. 

We observe that gain in vote-shares first increases with $p$, and then gradually decreases to $0$.  
In \cref{Heatmap-1}, a maximum gain of $17.85\%$ is observed when the budget ratio is $B_{A}/B_{B} = B_{A} = 0.3$.
In \cref{Heatmap-2}, gain is maximum ($5.75\%$) when controllers have equal budgets $B_{A} = B_{B} = 1$, and we note that difference in vote-shares is typically positively correlated with the budget $B_{A}$.
Finally in \cref{Heatmap-3}, we find that the gain in vote-shares reaches a maximum of $10.32\%$ at $B_{A}=0.6$.


Next, we examine the effect of competitor strategies on relative gain in vote-shares. We fix the available resources for both controllers at $B_{A} = B_{B} = N$ to reduce any impact of uneven budgets on the results. Once again we consider CP-Reg-High networks. 

We define competitor strategy for controller B, using $\epsilon_{B}$, which is the fraction of resources B allocates to nodes with negative edges. It is assumed that resources are always uniformly spread over targeted nodes, where $\epsilon_{B} \cdot B_{B}/p$ amount of influence is applied on every node that has a negative link, while $(1-\epsilon_{B}) \cdot B_{B}/(1-p)$ is given to the rest of the nodes (with strictly positive edges).
We gradually vary $0 \leq \epsilon_{B} \leq 1$. Here boundary conditions correspond to instances where controller B either strictly targets nodes with negative ties ($\epsilon_{B}=1$) or avoids them completely ($\epsilon_{B}=0$). Results are shown in \cref{Heatmap-4}. 

We observe that gain in vote-shares is high only when a fraction of the population has negative ties $p \leq 0.75$ and controller B targets these nodes (with negative ties) with less resources. As B discriminates between nodes with negative ties and the rest of the network, controller A stands to gain more from following a negative-tie aware approach.
The maximum gain of $10.13\%$ is obtained when controller B allocates a quarter of its budget $p_{B} = 0.25 \cdot B_{B}$ to all hub-nodes with negative edges where $p=0.5$.

\section{Analytical support}
\label{mean-field}
We now propose an analytical framework in support of our numerical results. Note that, obtaining closed-form analytical solution for \cref{optimisation} on networks with complex structures can be challenging. We simplify the problem by adopting a degree-based mean-field approach that approximates system dynamics and helps us obtain analytical expressions for optimal allocations. This approach works by grouping nodes with the same positive ($k_{a}$) and negative degrees ($k_{b}$). Here nodes are assigned $k_{a}$ and $k_{b}$ independently of each other, and there is no correlation between them. We then assume that the opinion state of nodes $x_{k_{a}k_{b}}$ is correlated to their overall degree ($\{k_{a},k_{b}\}$). In doing so, we homogenise the effect of a heterogeneous neighbourhood on the state of a node, which increases the  tractability of the problem. Additionally, we assume external controllers A and B uniformly target all nodes within a given class (with $a_{k_{a}k_{b}}$ and $ b_{k_{a}k_{b}}$ allocations respectively), which further reduces the degrees of freedom in our problem. 

We first derive the state of a node, with positive degree $k_{a}$ and negative degree $k_{b}$, from the steady-state equation \cref{steady-state} as
\begin{align}
    x_{k_{a}k_{b}} = \frac{a_{k_{a}k_{b}} + k_{b} + k_{a} \langle x_{a} \rangle - k_{b} \langle x_{b} \rangle}{k_{a} + k_{b} + a_{k_{a}k_{b}} + b_{k_{a}k_{b}}}.
    \label{ss}
\end{align}
Here $\langle x_{a} \rangle$ and $\langle x_{b} \rangle$ represent the expected state of a neighbouring node across a positive and a negative edge, respectively. 
Assuming that $P_{a} (k_{a})$  and $P_{b} (k_{b})$ describe the positive and negative degree distributions of a network, the expected behaviour of a neighbour at the end of a positive or a negative edge ($\langle x_{a} \rangle$ and $\langle x_{b} \rangle$) can be obtained as

\begin{align}
\begin{aligned}
     & \langle x_{a} \rangle = \sum_{k_{b}} P_{b} \sum_{k_{a}} P_{a} \frac{k_a}{\langle k_{a} \rangle} x_{k_{a}k_{b}} \\ 
     & \langle x_{b} \rangle = \sum_{k_{a}} P_{a} \sum_{k_{b}} P_{b} \frac{k_b}{\langle k_{b} \rangle} x_{k_{a}k_{b}}
\end{aligned}
\label{nbr}
\end{align}

Note that here the term $k_{a}/\langle k_{a} \rangle$ (or $k_{b}/\langle k_{b} \rangle$) ensures that a node with higher positive (or negative) degree $k_{a}$ (or $k_{b}$) has a higher chance of appearing as a neighbour exerting positive influence.

Now, using \cref{ss} and \cref{nbr} we obtain the following self-consistency relations,
\begin{align}
\begin{aligned}
     & \langle x_{a} \rangle = \langle \frac{k_{a}}{\langle k_{a} \rangle} \frac{a_{k_{a}k_{b}} + k_{b}}{\Delta} \rangle + \langle \frac{k_{a}^2}{\langle k_{a} \rangle} \frac{1}{\Delta} \rangle \langle x_{a} \rangle - \langle \frac{k_{a}k_{b}}{\langle k_{a} \rangle} \frac{1}{\Delta} \rangle \langle x_{b} \rangle 
     \\ 
     & \langle x_{b} \rangle = \langle \frac{k_{b}}{\langle k_{b} \rangle} \frac{a_{k_{a}k_{b}} + k_{b}}{\Delta} \rangle + \langle \frac{k_{a}k_{b}}{\langle k_{b} \rangle} \frac{1}{\Delta} \rangle \langle x_{a} \rangle - \langle \frac{k_{b}^2}{\langle k_{b} \rangle} \frac{1}{\Delta} \rangle \langle x_{b} \rangle
\end{aligned}
\end{align}
These can be further solved to finally arrive at the mean-field expressions 
\begin{align}
    \begin{split}
       \langle x_{a} \rangle &= \Bigg[\langle \frac{k_{a}}{\langle k_{a} \rangle} \frac{a_{ab}+k_{b}}{\Delta} \rangle - \frac{\langle \frac{k_{a}k_{b}}{\langle k_{a} \rangle} \frac{1}{\Delta}\rangle \langle \frac{k_{b}}{\langle k_{b} \rangle} \frac{a_{ab}+k_{b}}{\Delta} \rangle }{1 + \langle \frac{k_{b}^2}{\langle k_{b} \rangle} \frac{1}{\Delta} \rangle}\Bigg]
    \Bigg[1 - \langle \frac{k_{a}^2}{\langle k_{a} \rangle} \frac{1}{\Delta} \rangle + \frac{\langle \frac{k_{a}k_{b}}{\Delta} \rangle^2 \frac{1}{\langle k_{a} \rangle \langle k_{b} \rangle}}{1 + \langle \frac{k_{b}^2}{\langle k_{b}\rangle} \frac{1}{\Delta} \rangle}\Bigg]^{-1}, \\  
    \text{and,} \qquad \langle x_{b} \rangle  &= \Bigg[\langle \frac{k_{b}}{\langle k_{b}\rangle} \frac{a_{ab}+k_{b}}{\Delta} \rangle + \frac{\langle \frac{k_{b}}{\langle k_{b} \rangle} \frac{k_{a}}{\Delta}\rangle \langle \frac{k_{a}}{\langle k_{a} \rangle} \frac{a_{ab}+k_{b}}{\Delta} \rangle}{1 - \langle \frac{k_{a}^2}{\langle k_{a} \rangle} \frac{1}{\Delta} \rangle}\Bigg]
    \Bigg[1 + \langle \frac{k_{b}^2}{\langle k_{b} \rangle} \frac{1}{\Delta} \rangle + \frac{\langle \frac{k_{a}k_{b}}{\Delta} \rangle^2 \frac{1}{\langle k_{a} \rangle \langle k_{b} \rangle}}{1 - \langle \frac{k_{a}^2}{\langle k_{a} \rangle} \frac{1}{\Delta} \rangle}
    \Bigg]^{-1}.
    \end{split}
    \label{xa-xb}
\end{align}
Note that here $\Delta = a_{k_{a}k_{b}} + b_ {k_{a}k_{b}} + k_{a} + k_{b}$. 

Using the above, an approximation for the final vote-share at equilibrium can be obtained as
\begin{equation}
    X_{A} = \langle x_{k_{a},k_{b}} \rangle = \langle \frac{a_{ab}+k_{b}}{\Delta} \rangle + \langle \frac{k_{a}}{\Delta} \rangle \langle x_{a} \rangle - \langle \frac{k_{b}}{\Delta} \rangle \langle x_{b} \rangle,
    \label{XA-MF}
\end{equation}
where expressions for $\langle x_{a} \rangle$ and $\langle x_{b} \rangle$ come from \cref{xa-xb}. 

Using \cref{XA-MF} to derive closed-form analytical expressions for optimal allocations is still a complex task. Thus, we proceed with two alternative approaches. The first method uses a semi-analytical approach to obtain optimal allocations in simplified network structures using \cref{XA-MF}. 
Alternatively, we study the problem under limiting conditions, which further streamlines the analytical approach by adding more assumptions to the model. 

\subsection{Semi-analytical approach}
\label{semi-analytical}
We approximate optimal allocations by solving \cref{XA-MF} numerically in three types of networks, (i) Reg-Reg, (ii) CP-Reg-High and (iii) Reg-CP. 

In Reg-Reg networks, we have two types of nodes:   
(i) $p \cdot N$ nodes with $k_{a,1}=16$ positive edges and $k_{b,1}=4/p$ negative edges and the remaining (ii) $(1-p) \cdot N$ nodes with $k_{a,2}=16$ positive edges and no negative ties $k_{b,2}=0$. 
We assume controller A distributes $\epsilon_{A}$ fraction of its budget equally over all nodes with negative edges, $a_{k_{a,1}k_{b,1}} = \epsilon_{A} \cdot B_{A}/p$. The rest of the nodes receive $a_{k_{a,2}k_{b,2}} = (1-\epsilon_{A}) \cdot B_{A}/(1-p)$ allocations.

In CP-Reg-High networks, nodes can be segregated in two or three classes depending on the value of $p$. The positive component in these networks have a core-periphery structure with high-degree nodes $k_{a,1}=30$ and low-degree nodes $k_{a,2}=2$.
When $p=0.5$, we have two groups of nodes, (i) high-degree nodes with negative links and (ii) low degree nodes with only positive edges. As described earlier, resources are split into $\epsilon_{A}$ which is allocated to high-degree nodes and $(1-\epsilon_{A})$ given to low-degree nodes. For $p<0.5$, we have three groups of nodes, (i) high-degree nodes with negative links $(k_{a,1}+k_{b,1})$, (ii) high-degree nodes with only positive links $(k_{a,1})$ and (iii) low degree nodes with only positive links $(k_{a,2})$. The same can be derived for $p>0.5$, where all high-degree nodes and a fraction of low-degree nodes have negative links. In each case, the budget is split over three groups as $\epsilon_{A,1}$, $\epsilon_{A,2}$ and $[1 - ( \epsilon_{A,1}+\epsilon_{A,2})]$. We simultaneously solve for $\epsilon_{A,1}$ and $\epsilon_{A,2}$ to determine the optimal allocation on the network. The amount of budget allocated to nodes with negative edges is given by $\epsilon_{A} = \epsilon_{A,1}$ when $p<0.5$ and by $\epsilon_{A} = \epsilon_{A,1}+\epsilon_{A,2}$ when $p>0.5$.

\begin{figure}
  \begin{subfigure}[b]{0.475\textwidth}
    \includegraphics[width=\textwidth]{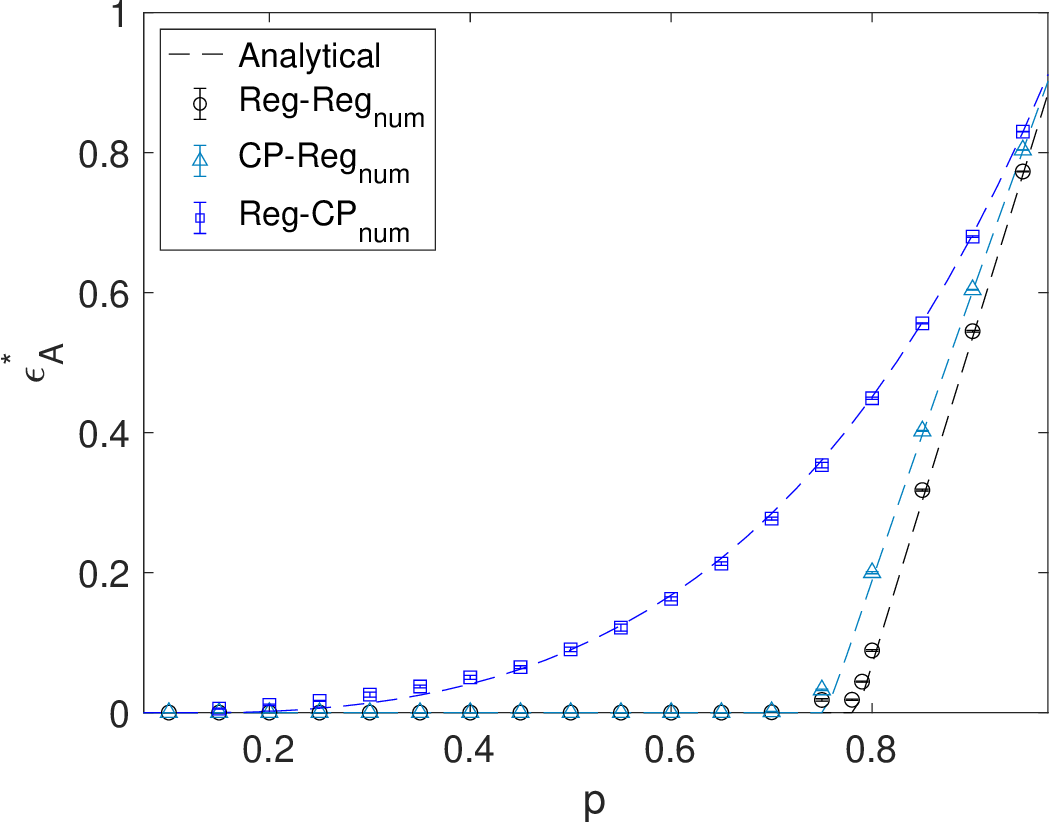}
    \caption{}
    \label{MF-1}
  \end{subfigure}
  \hspace{1em}
  \begin{subfigure}[b]{0.475\textwidth}
    \includegraphics[width=\textwidth]{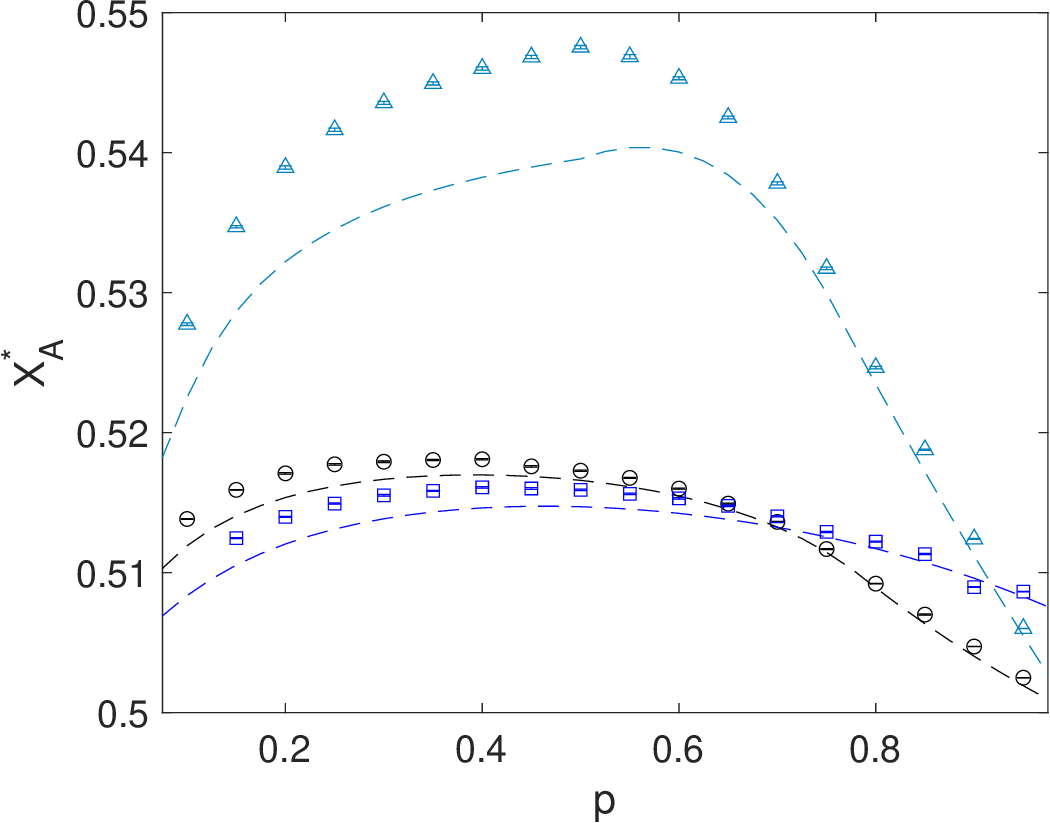}
    \caption{}
    \label{MF-2}
  \end{subfigure}
  \caption{Figures showing (a) the optimal allocations and (b) the vote-shares at equilibrium for values of $p \in [0.075,0.975]$. Results shown for three types of networks : (i) Reg-Reg, (ii) CP-Reg and (iii) Reg-CP. Analytical solutions are shown using dashed lines.
  Numerical results are obtained through simulations on networks of size N=1000 nodes and averaged over 10 networks. Error bars are shown for $95\%$ confidence intervals.}
  \label{MF}
\end{figure}

Optimal allocations are derived in Reg-CP networks using a comparable approach. Nodes form three clusters, (i) $p_{1}$ nodes with high-negative-degrees $k_{b,1} = 2(\langle k_{b} \rangle/p -1)$ and $k_{a,1}$ positive edges, (ii) $p_{2}$ nodes with low-negative-degree $k_{b,2} = 2$ and $k_{a,2}$ positive edges, and finally (iii) nodes with only $k_{a,3}$ positive edges. Here all classes of nodes have the same positive degree $k_{a,1} = k_{a,2} = k_{a,3} = 16$ and the fraction of nodes with negative edges are given as $p=p_{1}+p_{2}$. 
Budget allocation for each class of nodes is given as $a_{k_{a,1}k_{b,1}}=\epsilon_{A,1}/p_{1}$, $a_{k_{a,2}k_{b,2}}=\epsilon_{A,2}/p_{2}$ and $a_{k_{a,3}k_{b,3}}=(1 - \epsilon_{A})/(1-(p_{1}+p_{2}))$, where nodes with negative edges receive $\epsilon_{A} = \epsilon_{A,1}+\epsilon_{A,2}$ fraction of the budget.

For each type of network, we obtain related expressions using \cref{XA-MF}, which we solve semi-analytically to obtain optimal allocations $\epsilon_{A}^*$ and maximum vote-shares $X_{A}^*$. Results are presented in \cref{MF} against previous numerical results for comparison.

We compare our analytical results to numerical results obtained in the networks of size $N=1000$ nodes with $\langle k_{a} \rangle=16$ and $\langle k_{b} \rangle=4$, and are generated using the approach described in \cref{hetero}. Distribution of negative ties is controlled by varying $p$ between $0.075\leq p \leq 0.975$. In all cases, controller B targets all nodes equally $b_{k_{a,i}k_{b,i}} = 1$, $\forall i \in \{1,...,m\}$ where there are $m$ types (or groups) of nodes. 

Comparison between the semi-analytical and numerical results in \cref{MF-1,MF-2} shows that our numerical results are in good agreement with the theoretical solutions. There is, however, some discrepancy in vote-share results (\cref{MF-2}), which can be attributed to uniform allocations to each group of nodes in the analytical approach. The numerical method in contrast allows more sophisticated, flexible allocations to nodes of the same type, thereby realising higher vote-shares. 
In \cref{MF-1}, we note that for lower values of $p$, nodes with negative edges are strictly avoided in Reg-Reg and CP-Reg-High networks. However, as $p$ increases and negative edges are dispersed over the network, allocations to these nodes increase. This behaviour contrasts with Reg-CP graphs, where we find positive allocations to nodes with negative ties even at low values of $p$. We further observe that these allocations are to nodes with low-negative-degrees in the core-periphery negative components, hinting that allocations to a node may be a function of its negative degree.

\subsection{In the limit of large $\langle k_{a} \rangle$}
\label{limiting-case}
As an alternative to the semi-analytical approach presented in \cref{mean-field}, we expand the mean-field approximation in \cref{XA-MF} in the limit of large average positive degree, compared to controller budgets and average negative degree, i.e. $\langle k_{a} \rangle \gg \langle a_{k_{a}k_{b}} \rangle + \langle b_{k_{a}k_{b}} \rangle+\langle k_{b} \rangle$. This is a reasonable approximation as data for real-world networks have shown relatively small fractions of negative edges \cite{leskovec2010signed}. We begin by performing a series expansion on \cref{XA-MF} in the above limit to obtain,

\begin{equation}
\begin{split}
    X_{A} \approx \langle \frac{a_{k_{a}k_{b}}}{k_{a}} \rangle + \langle \frac{k_{b}}{k_{a}} \rangle + \frac{a_{k_{a}k_{b}}+\langle k_{b} \rangle}{a_{k_{a}k_{b}}+b_{k_{a}k_{b}}+2\langle k_{b} \rangle} - \frac{\langle \frac{(a_{k_{a}k_{b}}+k_{b})(a_{k_{a}k_{b}}+b_{k_{a}k_{b}}+2k_{b})}{k_{a}} \rangle}{a_{k_{a}k_{b}}+b_{k_{a}k_{b}}+2\langle k_{b} \rangle} \\ + \frac{a_{k_{a}k_{b}}+\langle k_{b} \rangle}{(a_{k_{a}k_{b}}+b_{k_{a}k_{b}}+2\langle k_{b} \rangle)^2}\Bigg( \langle \frac{k_{b}^2}{k_{a}} \rangle  + \langle \frac{(a_{k_{a}k_{b}}+b_{k_{a}k_{b}}+k_{b})(a_{k_{a}k_{b}}+b_{k_{a}k_{b}}+3k_{b})}{k_{a}} \rangle \\ 
    - \langle \frac{a_{k_{a}k_{b}}+b_{k_{a}k_{b}}+k_{b}}{k_{a}} \rangle \frac{a_{k_{a}k_{b}}+\langle k_{b} \rangle }{a_{k_{a}k_{b}}+b_{k_{a}k_{b}}+2\langle k_{b}\rangle} - \langle \frac{k_{b}}{k_{a}} \rangle  \frac{a_{k_{a}k_{b}}+\langle k_{b}\rangle}{a_{k_{a}k_{b}}+b_{k_{a}k_{b}}+\langle k_{b}\rangle} \Bigg).
\end{split}
\label{approx-Xa}
\end{equation}
We further use the above expression to derive optimal allocations $a_{k_{a}k_{b}}^*$, obtained as 

\begin{equation}
        a_{k_{a}k_{b}}^* = \frac{1}{2} \Bigg( \frac{\langle a_{k_{a}k_{b}}\rangle - \langle b_{k_{a}k_{b}}\rangle}{\langle b_{k_{a}k_{b}} \rangle +\langle k_{b} \rangle}  b_{k_{a}k_{b}} + \frac{\langle a_{k_{a}k_{b}} \rangle - 3\langle b_{k_{a}k_{b}} \rangle -2 \langle k_{b} \rangle}{\langle b_{k_{a}k_{b}} \rangle +\langle k_{b} \rangle} k_{b} + \langle a_{k_{a}k_{b}}\rangle + \langle b_{k_{a}k_{b}}\rangle+2\langle k_{b} \rangle \Bigg).
  \label{optimal-allo}
\end{equation}
Details of the derivation can be found in \cref{appendix-A}.

\begin{figure}
  \begin{subfigure}[b]{\textwidth}
  \centering
    \includegraphics[width=1\textwidth]{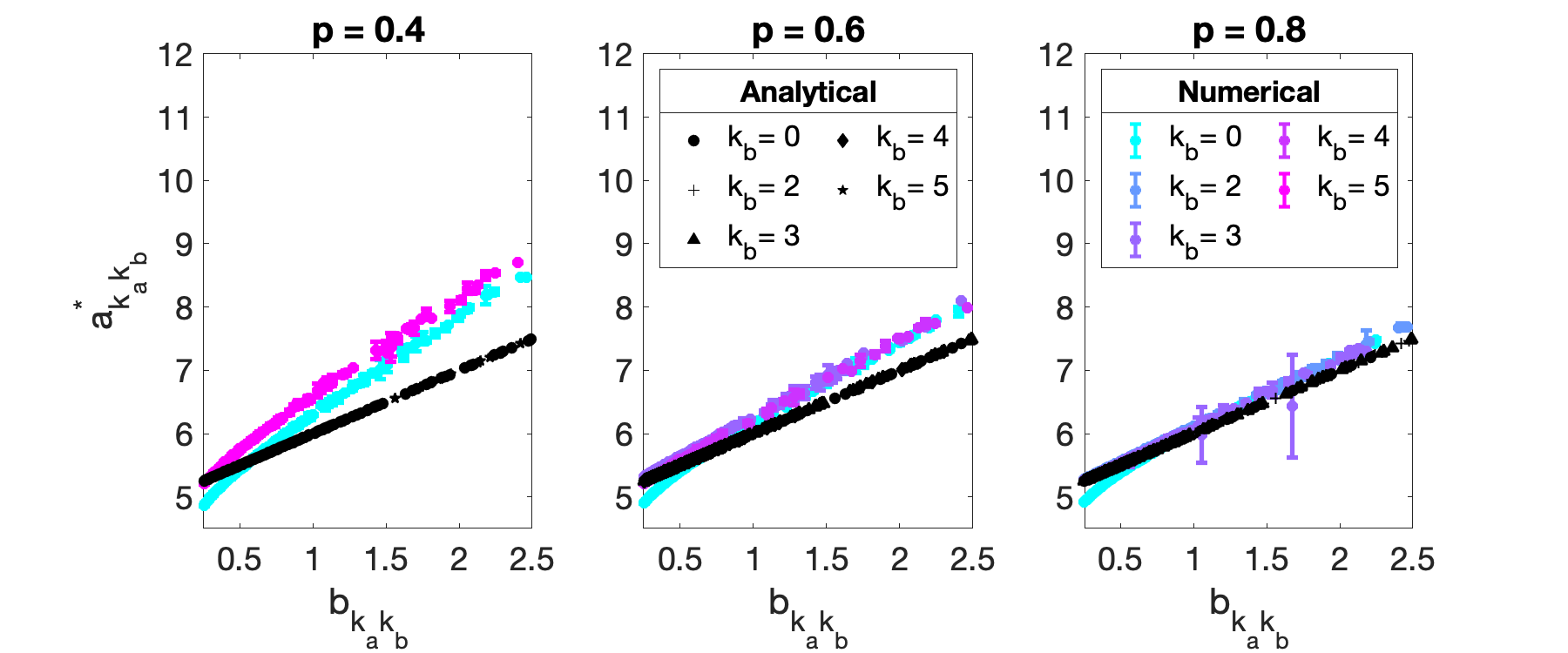}
    \caption{}
    \label{Ana-num-1}
  \end{subfigure}
  
    \begin{subfigure}[b]{\textwidth}
    \includegraphics[width=1\textwidth]{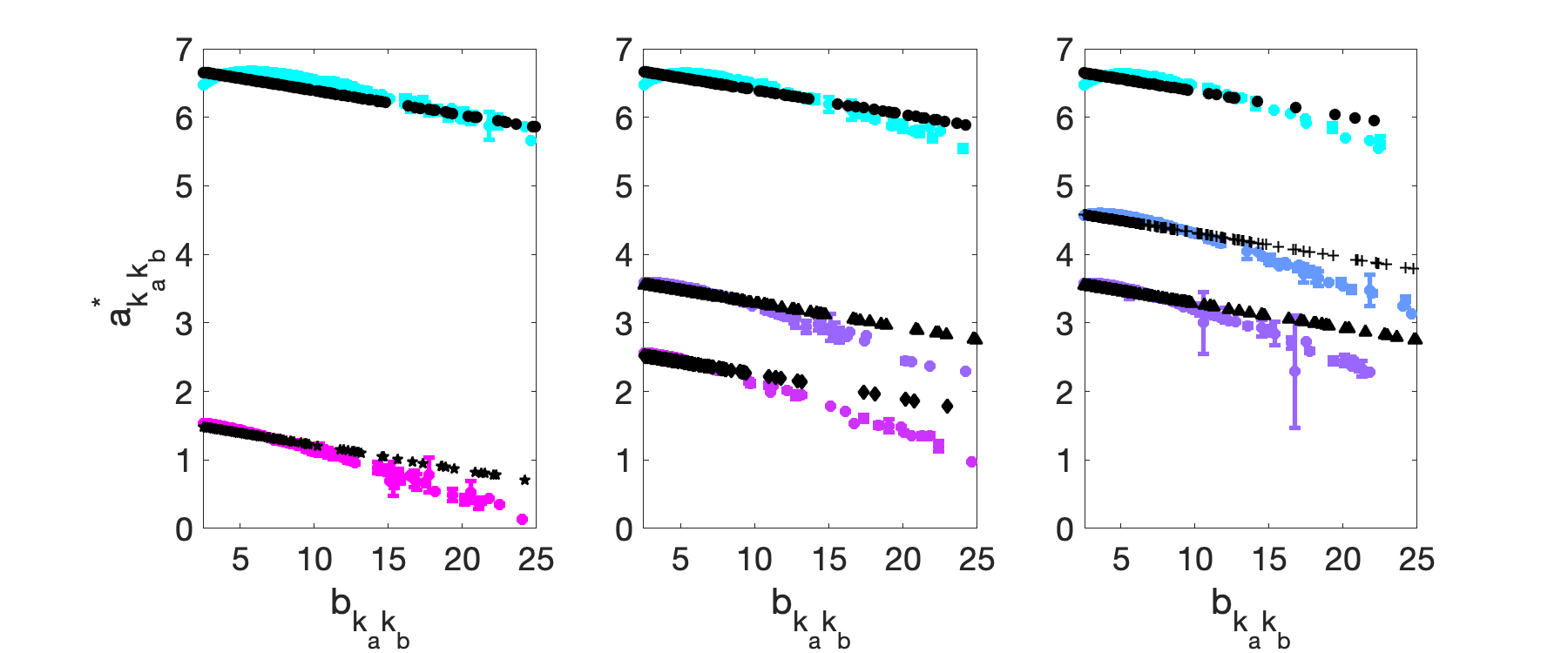}
    \caption{}
    \label{Ana-num-2}
  \end{subfigure}
  \caption{Figures showing mean allocations to nodes as a function of competitor allocations $b_{k_{a}k_{b}}$ in SF-Reg networks of size $N=1000$, where $\langle k_{a} \rangle = 50$ and $\langle k_{b} \rangle = 2$. We examine three instances of negative tie distributions $p \in [0.4,0.6,0.8]$ (left to right). Panel (a) shows positive correlations between optimal allocations and competitor allocations and Panel (b) shows negative correlations between optimal allocations and competitor allocations. Different symbols (and colours) correspond to nodes with different negative degrees $k_{b}$. Controllers have the following budgets: (a) $\langle a_{k_{a}k_{b}} \rangle = 5.5$ and $\langle b_{k_{a}k_{b}} \rangle = 0.5$, and (b) $\langle a_{k_{a}k_{b}} \rangle = 4.5$ and $\langle b_{k_{a}k_{b}} \rangle = 5$. Controller B here follows a $k_{a}-$dependent strategy. Numerical results are averaged over 10 networks. Error bars show 95\% confidence intervals.}
  \label{Ana-num-b}
  \end{figure}

From \cref{optimal-allo} we can see that the optimal allocation $a_{k_{a}k_{b}}^*$ to a node characteristically depends on its negative degree $k_{b}$ and the competitor influence $b_{k_{a}k_{b}}$ on it. We verify these findings in the remainder of the section using numerical simulations.  

We first validate the dependence of optimal allocations $a_{k_{a}k_{b}}^*$ on competitor allocations $b_{k_{a}k_{b}}$. SF-Reg networks are employed to test this relationship. Here the scale-free positive component is exploited to extract multiple data points for comparison between numerical and analytical results. For this reason we assume controller B has a $k_{a}$-dependent strategy. 
Note that there is a negative-degree dependent term in \cref{optimal-allo} that contributes to optimal allocations. We reduce the effect of this term in our analysis by first setting the per node budget for controller A to $\langle a_{k_{a}k_{b}} \rangle = 3\langle b_{k_{a}k_{b}} \rangle + 2\langle k_{b} \rangle$. However, this condition constrains the analysis to positive correlations between allocations from both controllers, as now $\langle a_{k_{a}k_{b}} \rangle > \langle b_{k_{a}k_{b}} \rangle$ (see the first term in \cref{optimal-allo}). When analysing the negative relation between optimal and competitor allocations, the budget is set such that $\langle a_{k_{a}k_{b}} \rangle < \langle b_{k_{a}k_{b}} \rangle$. In this case however, the negative-degree-dependent term contributes to the optimal allocation result. The regular structure of the negative component in the networks are chosen to limit the variation in optimal allocation $a_{k_{a}k_{b}}^*$ caused by negative degrees.

For our simulations we consider networks where $\langle k_{a} \rangle=50$, is sufficiently large compared to $\langle k_{b} \rangle=2$. Controller B distributes its budget in proportion to the positive degree of nodes ($p_{B,i} \propto k_{a,i}$). When analysing the positive relation between allocations $a_{k_{a}k_{b}}^*$ and $b_{k_{a}k_{b}}$, the budget for controllers A and B are $\langle a_{k_{a}k_{b}} \rangle=5.5$ and $\langle b_{k_{a}k_{b}} \rangle=0.5$ respectively. For negative correlations we choose $\langle a_{k_{a}k_{b}} \rangle=4.5$ and $\langle b_{k_{a}k_{b}} \rangle=5$ to avoid defying the positivity constraint on optimal allocations. We summarise results from the simulations in \cref{Ana-num-1,Ana-num-2}, where for the sake of brevity, we illustrate results in only three instances of $p \in [0.4,0.6,0.8]$.

\begin{figure} 
  \begin{subfigure}[b]{\textwidth}
    \includegraphics[width=1\textwidth]{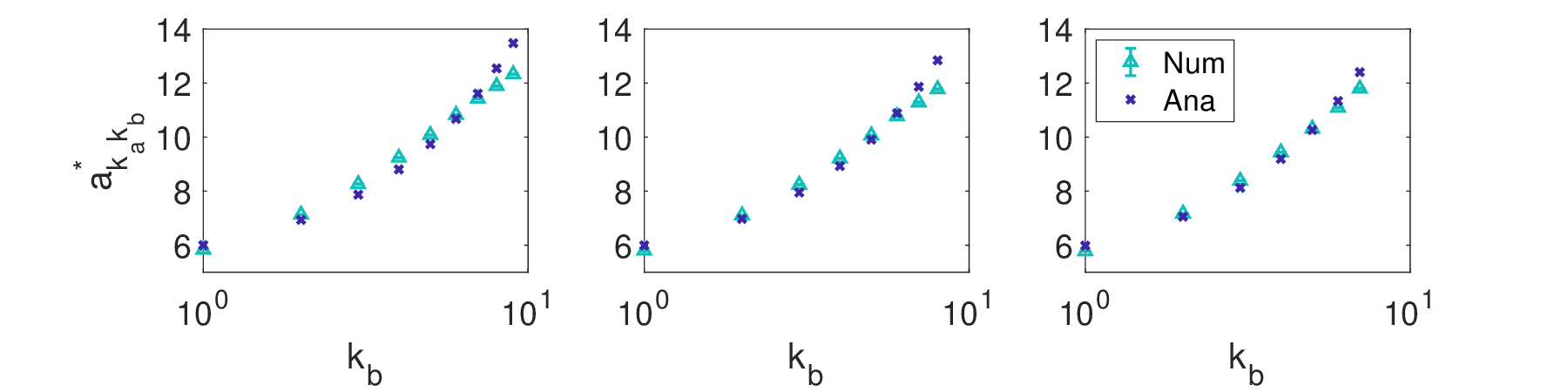}
    \caption{}
    \label{Ana-num-3}
  \end{subfigure}

  \begin{subfigure}[b]{\textwidth}
    \includegraphics[width=1\textwidth]{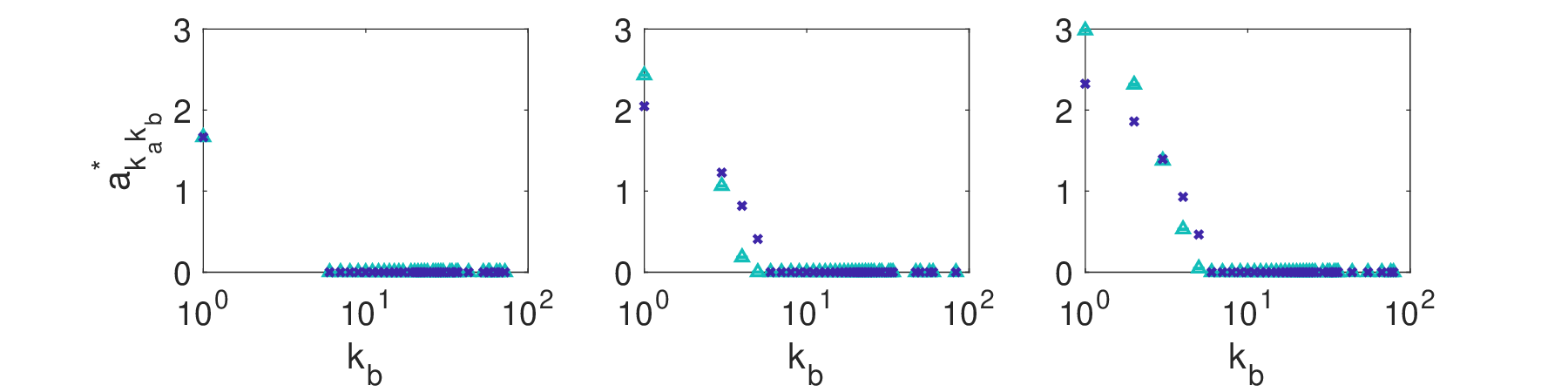}
    \caption{}
    \label{Ana-num-4}
  \end{subfigure}
  \caption{Figures showing mean allocations to nodes as a function of negative degrees $k_{b}$ in networks of size $N=1000$. We examine three instances of negative tie distributions $p \in [0.4,0.6,0.8]$ (left to right). Figure (a) illustrates positive correlations between optimal allocations and negative degrees $k_{b}$ in Reg-ER networks where $\langle k_{a} \rangle = 50$ and $\langle k_{b} \rangle = 2$. Similarly, Figure (b) shows a negative relation between optimal allocations and negative degrees in Reg-SF networks where $\langle k_{a} \rangle = 16$ and $\langle k_{b} \rangle = 4$.
  Controllers have the following budgets: (a) $\langle a_{k_{a}k_{b}} \rangle = 7$ and $\langle b_{k_{a}k_{b}} \rangle = 0.5$, and (b) $\langle a_{k_{a}k_{b}} \rangle = 1$ and $\langle b_{k_{a}k_{b}} \rangle = 1$. Controller B targets the network uniformly. Numerical results are averaged over 10 networks. Error bars show 95\% confidence intervals.}
  \label{Ana-num-kb}
\end{figure}

Next, we explore the dependence of optimal allocations $a_{k_{a}k_{b}}^*$ on the negative degree $k_{b}$ of nodes. For the same reason as before, we now study positive regular graphs with negative heterogeneous components. Note that the correlation is positive when $ \langle a_{k_{a}k_{b}} \rangle > 3 \langle b_{k_{a}k_{b}} \rangle + 2\langle k_{b} \rangle$, and negative when $\langle a_{k_{a}k_{b}} \rangle < 3 \langle b_{k_{a}k_{b}} \rangle  - 2\langle k_{b} \rangle $. We conduct simulations to test each relation and present the results in \cref{Ana-num-3,Ana-num-4}.
To analyse the positive correlation, we run simulations in Reg-ER networks where $\langle k_{a} \rangle =50$ and $\langle k_{b} \rangle =2$. Here controller B evenly distributes $B_{B} = \langle b_{k_{a}k_{b}} \rangle \cdot N = 0.5 \cdot N$ resources to the network. Uniform allocation from the adversary limits the effect of the corresponding $b_{k_{a}k_{b}}$ term on optimal allocations. We assume controller A has a budget of $\langle a_{k_{a}k_{b}} \rangle = 7$ which satisfies the necessary condition $\langle a_{k_{a}k_{b}} \rangle > 3 \langle b_{k_{a}k_{b}} \rangle  - 2\langle k_{b} \rangle $. 
For the second sub-case, we use a Reg-SF network with $\langle k_{a} \rangle =16$ and $\langle k_{b} \rangle =4$. Controllers have equal budgets $\langle a_{k_{a}k_{b}} \rangle = \langle b_{k_{a}k_{b}} \rangle =1$, which reduces the numeric value of the $b_{k_{a}k_{b}}$-dependent term in \cref{optimal-allo} to $0$. As before, controller B targets the network uniformly. 

As shown in \cref{Ana-num-b,Ana-num-kb}, we find that analytical results approximate numerical simulations reasonably well in the limit of large average positive degrees. 
\cref{Ana-num-1} corroborates a positive linear relation between optimal allocations $a_{k_{a}k_{b}}^*$ and competitor allocation $b_{k_{a}k_{b}}$ under conditions of excess budget. We also observe that agreeability between numerical and analytical results increases with homogeneity of negative edges in the network ($p=0.8$). Furthermore, as expected, we find that negative degree of the node does not influence optimal allocations (as $\langle a_{k_{a}k_{b}} \rangle  = 3 \langle b_{k_{a}k_{b}} \rangle + 2\langle k_{b} \rangle$). \cref{Ana-num-2} verifies a negative relation between $a_{k_{a}k_{b}}^*$ and $b_{k_{a}k_{b}}$ when resources are scarce. We find that optimal allocations now depend on the negative degree of the node.

In \cref{Ana-num-2,Ana-num-3}, we measure correlations between optimal allocations and negative degree of nodes $k_{b}$. \cref{Ana-num-2} shows a positive linear relation with negative degree $k_{b}$ for larger budgets $ \langle a_{k_{a}k_{b}} \rangle > 3 \langle b_{k_{a}k_{b}} \rangle + 2\langle k_{b} \rangle$. 
Contrarily, an inverse relationship is observed with $k_{b}$, in \cref{Ana-num-3}, when budget $\langle a_{k_{a}k_{b}} \rangle < 3 \langle b_{k_{a}k_{b}} \rangle  - 2\langle k_{b} \rangle$.

The above results show that the proposed analytical framework provides a good estimate of the system dynamics in heterogeneous signed networks. We further show that in the limit of large average positive degree, optimal allocations depend on negative degree and competitor allocations on nodes, and not on the positive degree. The type of correlation, positive or negative, relies on the budget available to the focal controller. When resources are in excess, allocations increase linearly with competitor allocation and negative degree, whereas under low budget conditions this relation becomes inversely proportional.

\section{Game-theoretic setting}
\label{game-theory}
So far we have optimised allocations for controller A against a passive adversary B. In this section, we explore the problem in a game-theoretic setting where both controllers A and B actively optimise budget distributions over the network. In our game, controllers represent players who optimise their respective strategies ($p_{A}$ and $p_{B}$), to maximise their utilities or vote-shares ($X_{A}$ and $X_{B}$). 

Consider a setting where two companies use word-of-mouth marketing on a social network. Each company has a budget which they distribute over the network to maximise adoption of their product.
One may assume that neither company has any prior knowledge of the competitor's allocation on the network, as there is no incentive to disclose this information in advance. However, companies can gain this information through their interaction with the network during the marketing process and they can then use this knowledge to strategise in following campaigns.  
In our model this translates to a scenario where controllers simultaneously and iteratively target the network. Each player observes the opponent's strategy in the ongoing game and then plays the best response to this strategy in the following round. In our model, allocation vectors are revealed at the end of each round, which then helps players strategise in the next round.
Finally, we assume multiple iterations of the game are played until a state is achieved from which neither controller deviates. In the real world, this would correspond to a stable state where companies obtain their optimal outcome and are no longer motivated to change their marketing strategies.  

In the following, we aim to explore how much a controller can gain from the knowledge of negative ties. We address this by comparing outcomes in two scenarios. First, we consider a setting in which the controller A has complete knowledge of the network structure (with full knowledge of negative ties), while controller B is unaware of the negative edges. Second, both controllers A and B observe the entire network as an unsigned (positive) graph. We then measure how much the controller A benefits from the knowledge of the network structure by comparing both scenarios. 



\begin{figure}
 \centering
  \begin{subfigure}[b]{0.85\textwidth}
    \includegraphics[width=\textwidth]{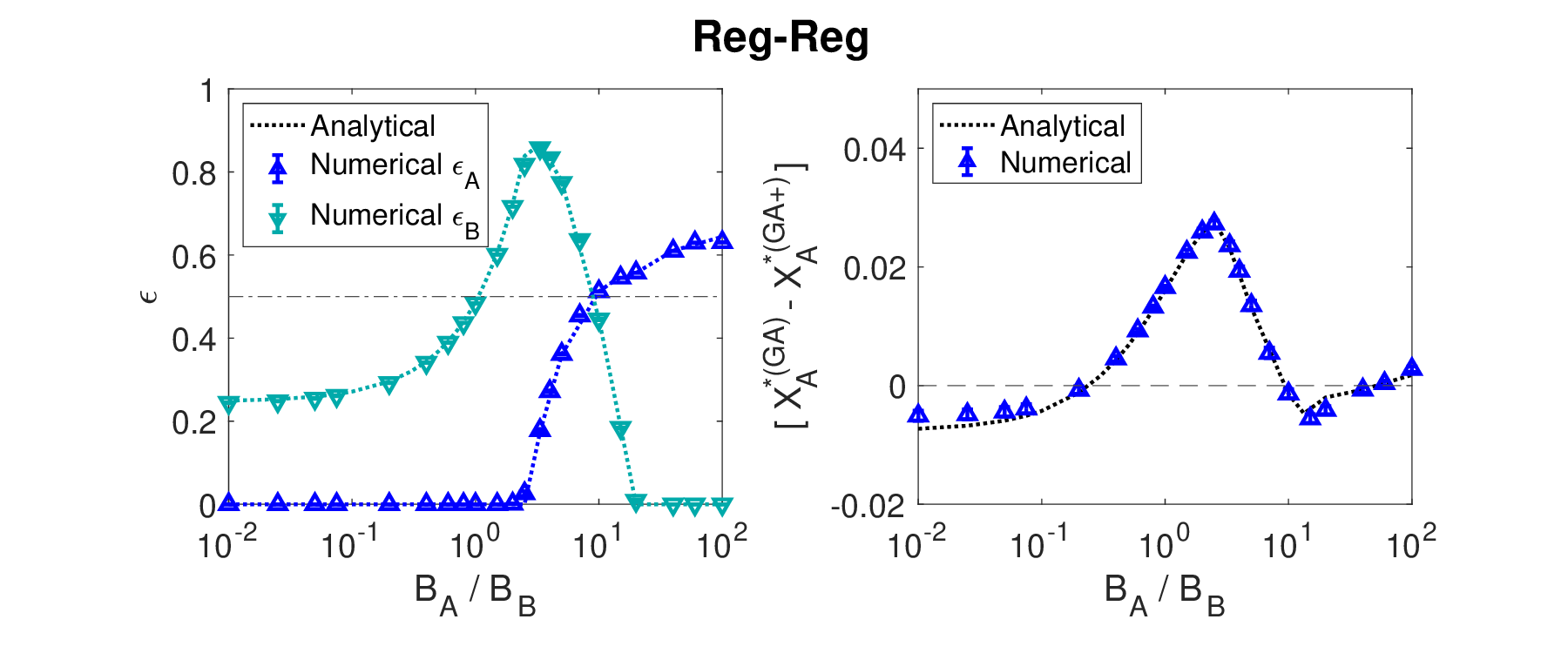}
    \caption{}
    \label{GT-1}
  \end{subfigure}
    \begin{subfigure}[b]{0.85\textwidth}
    \centering
    \includegraphics[width=\textwidth]{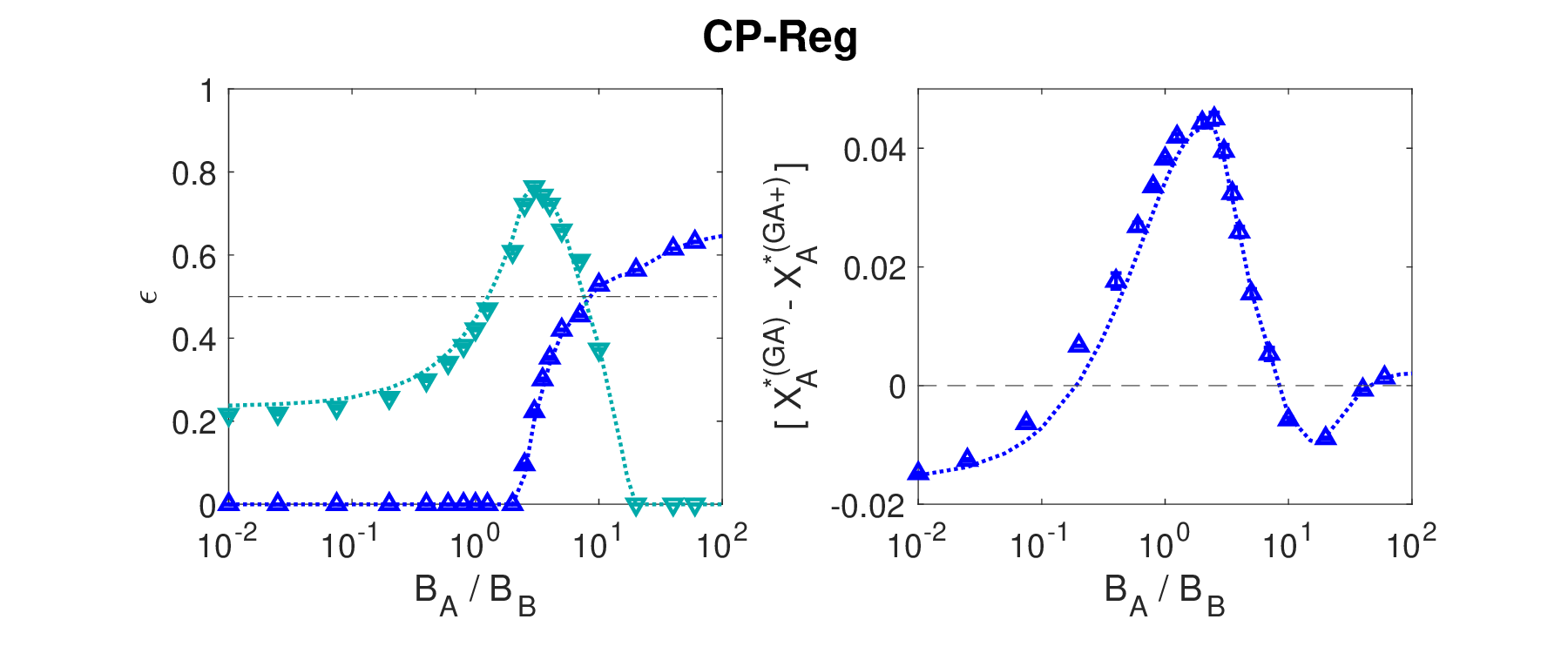}
    \caption{}
    \label{GT-2}
  \end{subfigure}
      \begin{subfigure}[b]{0.85\textwidth}
    \centering
    \includegraphics[width=\textwidth]{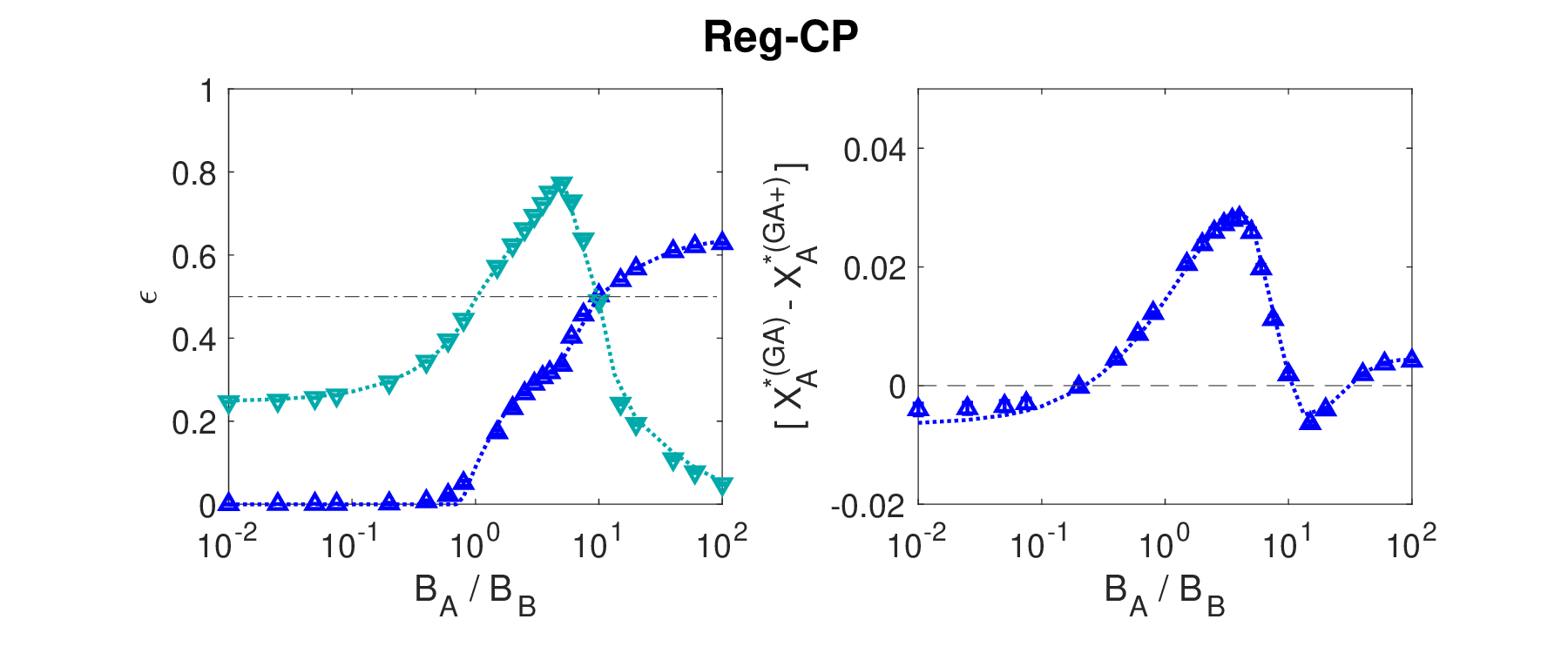}
    \caption{}
    \label{GT-3}
  \end{subfigure}
  \caption{Figures compare analytical and numerical results of equilibrium states. Figures on the left depict change in strategies ($\epsilon$) for both players as the budget ratio changes. Here $\epsilon$ is the fraction of resource given to nodes with negative edges. Figures on the right show gain in true vote-shares at equilibrium $[X_{A}^{*(GA)} - X_{A}^{*(GA+)}]$. Simulations are conducted in networks of size N=200,  $\langle k_{a} \rangle = 16$, $\langle k_{b} \rangle = 4$ where negative ties are distributed only over half the network $(p=0.5)$. 
  We consider three network types (a) Reg-Reg, (b) CP-Reg-High and (c) Reg-CP networks. In Reg-CP networks, $\epsilon=\epsilon_{1}+\epsilon_{2}$, where $\epsilon_{1}$ and $\epsilon_{2}$ are fractions of resources given to high-negative degree and low-negative degree nodes in the negative core-periphery component respectively. Numerical simulations are run with a step-size $\eta=5$ and terminated using a $\mu=10^{-7}$ approximation factor. Results are averaged over 10 networks and show error bars with $95\%$ confidence intervals. }
  \label{GT}
\end{figure}

We first solve the problem semi-analytically and then compare it to numerical results in \cref{GT}.  
To begin, we derive an analytical expression for the instance where controllers do not have complete knowledge of the network. Say the network is a signed network where a node $i$ has $k_{a}$ positive edges and $k_{b}$ negative edges. As controllers cannot discriminate between positive and negative edges, they assume that the node $i$ has degree $k_{i}$ where $k_{i}=k_{a,i}+k_{b,i}$. 

Using the degree-based mean-field approach discussed in \cref{mean-field} we assume that node behaviour $\langle x_{k} \rangle$ strongly correlates with their degree $k$. Now, using the steady-state equation (\cref{steady-state}) we obtain the state of a node with degree $k$ as,

\begin{align}
    x_{k} = \frac{k\langle x \rangle+a_{k}}{k+a_{k}+b_{k}},
    \label{x_k}
\end{align} 
where $a_{k}$ and $b_{k}$ are allocations from controllers A and B for the given class of nodes, and $\langle x \rangle$ is the average behaviour of a node in the network. Here controllers are assumed to distribute allocations uniformly within each class. 

Given a degree distribution $P_{k}$, the expected behaviour of a node in the network is,
\begin{align}
    \langle x \rangle = \frac{\sum_{k} P_{k} k x_{k}}{\langle k \rangle} = \frac{\sum_{k} P_{k} k x_{k}}{\sum_{k} P_{k} k}.
    \label{<x>}
\end{align}

Using \cref{x_k} and \cref{<x>}, we obtain 
\begin{align}
    \langle x \rangle  = \frac{\sum_{k} \frac{P_{k} k a_{k}}{k+a_{k}+b_{k}}}{\sum_{k} \frac{P_{k} k (a_{k}+b_{k})}{k+a_{k}+b_{k}}}.
\end{align}

This leads us to an expression for the total vote-share at equilibrium,

\begin{align}
    X_{A}^{(+)} = \sum_{k} P_{k}x_{k} = \frac{(\sum_{k} \frac{P_{k} k }{k+a_{k}+b_{k}})(\sum_{k} \frac{P_{k} k a_{k}}{k+a_{k}+b_{k}})}{(\sum_{k} \frac{P_{k} k (a_{k}+b_{k})}{k+a_{k}+b_{k}})}+ (\sum_{k} \frac{P_{k} k}{k+a_{k}+b_{k}}).
    \label{XA-positive}
\end{align}

Although we show the above derivations for A, the same process can be repeated to determine the utility function for controller B,
\begin{align}
    X_{B}^{(+)} = \frac{(\sum_{k} \frac{P_{k} k }{k+a_{k}+b_{k}})(\sum_{k} \frac{P_{k} k b_{k}}{k+a_{k}+b_{k}})}{(\sum_{k} \frac{P_{k} k (a_{k}+b_{k})}{k+a_{k}+b_{k}})}+ (\sum_{k} \frac{P_{k} k}{k+a_{k}+b_{k}}).
    \label{XB-positive}
\end{align}

Note that \cref{XA-positive} is the utility function we use for controller A when they cannot observe negative edges, while \cref{XA-MF} guides a negative-tie sensitive approach.
 To determine how much controller A gains from exploiting their knowledge of the network in the game-theoretic setting, we compare vote-shares obtained using both approaches.
 We first determine the optimal strategies at equilibrium, once using \cref{XA-MF} and then \cref{XA-positive}, both against a negative-tie agnostic competitor B whose utility is given by \cref{XB-positive}. In the case where both controllers are unaware of negative ties in the network, the utility functions are simultaneously solved to arrive at the equilibrium state $[X_{A},X_{B}]$. The corresponding strategies at this point are given by $[p_{A},p_{B}]$. 
 Given that both controllers are negative-tie agnostic, they observe their utilities at $[p_{A},p_{B}]$ as $[X_{A},X_{B}]$, which is different to their actual utilities $[\widehat{X}_{A}^{*},\widehat{X}_{B}^{*}]$ which should include negative-tie dynamics and is calculated by replacing $p_{A}$ and $p_{B}$ in \cref{optimisation}. In the other instance, controller A has access to their actual utility function at all times. Once an equilibrium state is reached, $[p_{A}^*,p_{B}^*]$ are used the determine the actual utility for B $\widehat{X}_{B}^{*}]$. 
 Here it is important to note that both controllers assume the network structure to be common knowledge at all times. Therefore, from both their perspectives, the game always appears to be a constant sum game.
 
 While solving the problem semi-analytically, we express strategies using $\epsilon_{A}$ (or $\epsilon_{B}$), which represent the fraction of resources allocated by a controller to nodes with negative ties. Solving $X_{A}$ and $X_{B}$ simultaneously, we obtain the equilibrium state $(\epsilon_{A}^{*},\epsilon_{B}^{*})$, and finally measure the difference in vote-shares. 
 For our simulations, we explore three types of networks that have been considered so far: (i) Reg-Reg, (ii) CP-Reg-High and (iii) Reg-CP, where $\langle k_{a} \rangle=16$ and $\langle k_{b} \rangle=4$. For brevity, we examine only one instance of negative tie distribution with $p=0.5$. 
Note that for Reg-Reg and CP-Reg-High networks, resources are split between two node classes as $\epsilon_{A}$ and $1-\epsilon_{A}$. The same is true for controller B. For Reg-Reg networks these are (i) $\{k_{a,1},k_{b,1}\}=\{16,8\}$ and (ii) $\{k_{a,2},k_{b,2}\}=\{16,0\}$, and for CP-Reg-High networks (i) $\{k_{a,1},k_{b,1}\}=\{30,8\}$ and (ii) $\{k_{a,2},k_{b,2}\}=\{2,0\}$. In Reg-CP networks, we have three classes of nodes (i) for $p_{1}=0.25$, $\{k_{a,1},k_{b,1}\}=\{16,14\}$, (ii) for $p_{2}=0.25$, $\{k_{a,2},k_{b,2}\}=\{16,2\}$ and finally (iii) for $p_{3}=0.5$, $\{k_{a,3},k_{b,3}\}=\{16,0\}$. Consequently the budget is split three-ways into $\epsilon_{1}$, $\epsilon_{2}$ and $\epsilon_{3} = 1 - (\epsilon_{1}+\epsilon_{2})$.  
 In all cases, the necessary boundary conditions are imposed. 
 \cref{GT} shows how equilibrium strategies and gain in vote-shares depend on budget ratios. Analytical results are complemented with numerical simulations. $GA$ is used as an optimiser when controller A has complete knowledge of the network and conversely $GA^{(+)}$ is used when they cannot discern polarity of the edges. In both cases, controller B uses $GA^{(+)}$ to determine best response. 
 
We find that simulations always converge to an equilibrium in all cases.
When both controllers are blind, the equilibrium strategy is identical for both controllers which is to uniformly target the network ($\epsilon_{A} = \epsilon_{B} = p = 0.5$) \cite{romero2021shadowing}. On the other hand, controller A avoids nodes with negative ties when they have full knowledge of the network and only have limited resources. This strategy influences the allocation patterns of competitor B, who is driven by the competitor's allocation to nodes and thus diverts less resources to these nodes. We see that by avoiding negative edges in the network, controller A inadvertently gives out information about negative ties in the network \textemdash through their own prioritisation of nodes, which then results in a loss in vote-shares as shown in \cref{GT}. 

 As the budget ratio increases, controller A starts redirecting some resources to nodes with negative edges. When this occurs, controller B increases their allocations to these nodes. As the budget ratio increases to $B_{A}/B_{B} \approx 5$, controller B stops competing over nodes with negative ties and begins to gradually reduce their allocation ($\epsilon_{B}$) to them. Around $B_{A}/B_{B} = 10$, both controllers target the network uniformly ($\epsilon_{A} = \epsilon_{B} = p = 0.5$), and beyond this point, $\epsilon_{B}$ continues to decrease as $\epsilon_{A}$ increases steadily. Once again, we find that controller A loses vote-shares by using their knowledge of negative edges in the network, as they start pursuing nodes with negative edges with more than half of their budget $\epsilon_{A} > 0.5$. However, as budget ratio increases to $B_{A}/B_{B} \approx 10^2$, controller A once again starts gaining from the knowledge of the network structure.
 
We also note that the results are similar across all three types of networks, and numerical simulations closely approximate semi-analytical results in all cases. Gain in vote-shares is significant in CP-Reg networks compared to the other networks. We find early allocations to nodes with negative edges in Reg-CP networks, which is consistent with results in \cref{MF}. Moreover, we observe that controller B never completely avoids negative edges in the Reg-CP networks. 

%% file: 4-discussion.tex
\section{Discussion}
\label{Summary}
The study of opinion dynamics has been conventionally done on networks with strictly positive edges. In the real world however, networks often contain negative social connections, which can spread negative or opposing influence, thus creating a need to understand how these edges affect influence maximisation efforts in networks. 
To address this concern, we present a model for competitive spread of opinions in signed networks under voter dynamics. For comparison, we propose a complementary approach where controllers only observe the absolute weights of all edges i.e. they consider all edges to be positive. In both instances we present gradient ascent algorithms to numerically solve the problem in large-scale arbitrary networks. We test the robustness of our results in networks of varied structures under diverse budget conditions and adversarial allocations. 
We find that in networks where 20\% of edges are negative, controllers gain maximally (nearly 18\%) from awareness of negative edges, under conditions of scarce resources, and against competitors who deliberately avoids nodes with negative connections. 

We also propose a supporting theoretical approach to verify the accuracy of our algorithms. We present closed-form solutions in simplified network structures that provide further insights to the problem. We observe that in networks with highly concentrated positive links, allocations on nodes are driven by their negative degrees and the competitor's allocation on these nodes. Finally, we examine the problem under game-theoretic settings, where we highlight conditions under which a controller could lose vote-shares by implementing strategies that use the knowledge of negative ties in the network. Specifically, we show that when controllers have considerably less resources (or in some cases, excess budget), their prioritisation of nodes to target, may inadvertently disclose knowledge of negative ties to a competitor who was otherwise unaware, thus compromising their position of advantage.

The results in this paper present compelling evidence for considering negative ties in any influence maximisation exercise and thus contributes to the literature on competitive opinion dynamics in signed networks. Possible extensions to this work could include studying the problem under different constraint functions. For instance, the effect of modified budget constraints that explore the implications of an additional cost to retrieve information about the presence of negative ties, on influence maximisation efforts.
Additionally, this problem could be further studied in other realistic opinion models (e.g. Deffaunt model). 


%% file: acknowledgments.tex

\paragraph*{Funding Statement.} This  research  was  sponsored  by  the  U.S.  Army  Research Laboratory  and  the  U.K.  Ministry  of  Defence  under  Agreement Number W911NF-16-3-0001. The views and conclusions contained in this document are those of the authors and should not  be  interpreted  as  representing  the  official  policies,  either expressed or implied, of the U.S. Army Research Laboratory, the  U.S.  Government,  the  U.K.  Ministry  of  Defence  or  the U.K.  Government.  The  U.S.  and  U.K.  Governments  are  authorized  to  reproduce  and  distribute  reprints  for  Government purposes notwithstanding any copyright notation hereon.

%% file: A-appendix.tex
\section*{Appendix}
\appendix
\numberwithin{equation}{section}

\section{Removing negative ties}
\label{appendix-rem}
When controllers are unable to detect or observe negative edges in the network, i.e. $w_{ij}=max(0,w_{ij})$, the optimisation problem reduces to
\begin{align}
    p_{A}^* &= \text{argmax}_{p_{A}} X_{A}^{*(\phi)}(L^{(\phi)},p_{B},B_{A}).
\end{align}

where $L^{(\phi)}$ is the updated Laplacian. 
Following the same process as before we use the gradient $\nabla_{p_{A}} X_{A}^{(\phi)} = 1/N \Vec{1}^{T} [L^{(\phi)}+diag(p_{A}+p_{B})]^{-1}(I - diag(x_{A}^{(\phi)})$ to optimise allocations $p_{A}^*$ in a gradient ascent algorithm $GA^{(\phi)}$. Here $    x_{A,i}^{*(\phi)} = (p_{A,i} + \sum\limits_{j}^{k_a}w_{ji} x_{A,j}^{(\phi)}) / (\sum\limits_{j}^{k_a}w_{ji} + p_{A,i} + p_{B,i})$.

\begin{figure}
  \centering
    \includegraphics[width=0.7\textwidth]{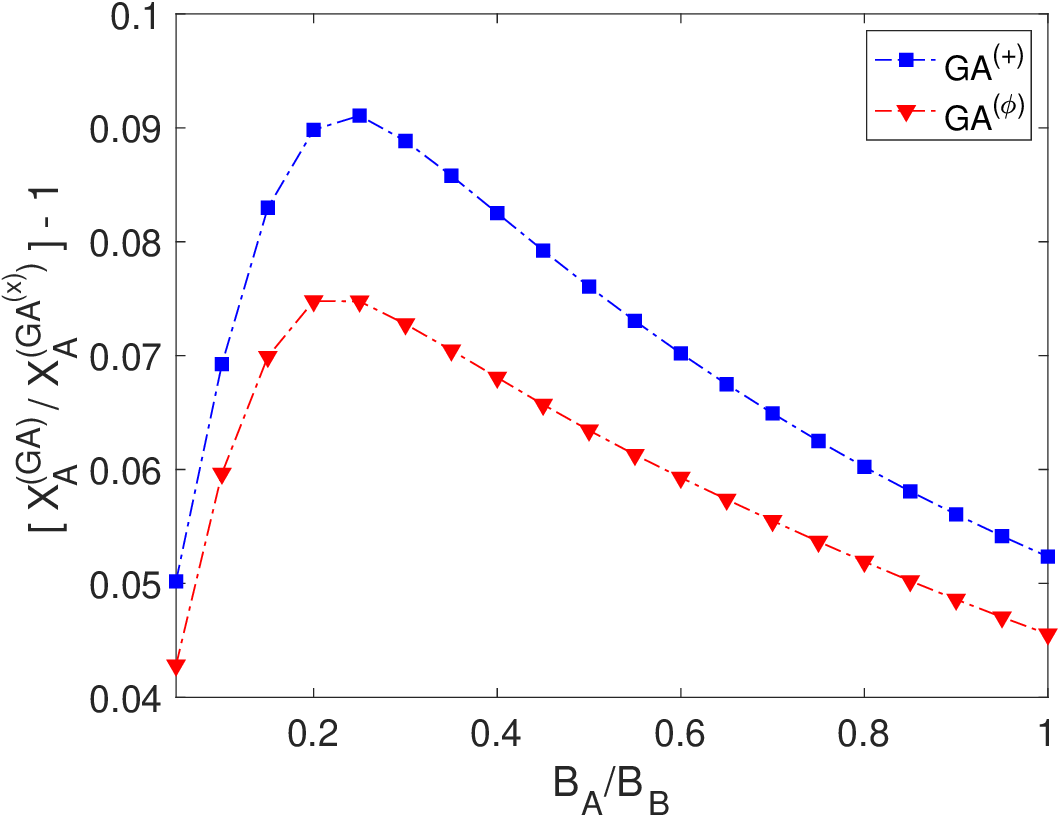}
    \caption{Figure showing gain in vote-shares when comparing the negative-tie sensitive optimisation approach $GA$, to traditional approaches, $GA^{(+)}$ and $GA^{(\phi)}$ for budget ratios $B_{A}/B_{B} \in [0.05,1]$. Controller B here targets the network uniformly.}
    \label{rem}
  \end{figure}

We then run $GA$, $GA^{(+)}$ and $GA^{(\phi)}$ on the Bitcoin network and present the respective gains in vote-shares in \cref{rem}.

We find that the method assuming negative edges in the network to be positive $GA^{(+)}$ consistently outperforms $GA^{(\phi)}$, where negative edges are not considered at all. 
To further show that the vote-shares obtained through both methods are identical in undirected networks (and that comparing our results to only $GA^{(+)}$ is sufficient), we look at the allocation expression in \cref{pos-allo}. Here we find that the optimal allocation, in the absence of any knowledge of negative edges, depends solely on the individual budgets of the controllers and the adversarial allocations on the nodes, but not on the degrees of the nodes.   

\section{Limiting case}
\label{appendix-A}
We begin with the series expansion of the steady-state equation in \cref{XA-MF} to obtain,
\begin{align}
    X_{A} = \langle \frac{a_{k_{a}k_{b}}}{k_{a}} \rangle + \langle \frac{k_{b}}{k_{a}} \rangle + \langle 1 - \frac{a_{k_{a}k_{b}}+b_{k_{a}k_{b}}+k_{b}}{k_{a}} \rangle  \langle x_{a} \rangle - \langle \frac{k_{b}}{k_{a}} \rangle \langle x_{b} \rangle,
    \label{x-exp}
\end{align}
where a second-order expansion for $\langle x_{a} \rangle$ gives us, 
\begin{align}
    \langle x_{a} \rangle = \frac{a+\langle k_{b} \rangle - \langle \frac{(a+k_{b})(a+b+2k_{b})}{k_{a}} \rangle }{a+b+2\langle k_{b} \rangle} + \frac{a+\langle k_{b} \rangle }{(a+b+2k_{b})^2}\Bigg( \langle \frac{k_{b}^2}{k_{a}} \rangle + \langle \frac{(a+b+k_{b})(a+b+3k_{b})}{k_{a}} \rangle \Bigg),
    \label{xa-exp}
\end{align}
and a zero-order expansion for $\langle x_{b} \rangle$ gives us
\begin{align}
  \langle x_{b} \rangle = \frac{a+\langle k_{b} \rangle}{a+b+\langle k_{b} \rangle}.
  \label{xb-exp}
\end{align}

Finally, replacing \cref{xa-exp,xb-exp} in \cref{x-exp} and ignoring higher-order terms, we obtain
\begin{equation}
\begin{split}
    X_{A} \approx \langle \frac{a_{k_{a}k_{b}}}{k_{a}} \rangle + \langle \frac{k_{b}}{k_{a}} \rangle + \frac{a_{k_{a}k_{b}}+\langle k_{b} \rangle}{a_{k_{a}k_{b}}+b_{k_{a}k_{b}}+2\langle k_{b} \rangle} - \frac{\langle \frac{(a_{k_{a}k_{b}}+k_{b})(a_{k_{a}k_{b}}+b_{k_{a}k_{b}}+2k_{b})}{k_{a}} \rangle}{a_{k_{a}k_{b}}+b_{k_{a}k_{b}}+2\langle k_{b} \rangle} \\ + \frac{a_{k_{a}k_{b}}+\langle k_{b} \rangle}{(a_{k_{a}k_{b}}+b_{k_{a}k_{b}}+2\langle k_{b} \rangle)^2}\Bigg( \langle \frac{k_{b}^2}{k_{a}} \rangle  + \langle \frac{(a_{k_{a}k_{b}}+b_{k_{a}k_{b}}+k_{b})(a_{k_{a}k_{b}}+b_{k_{a}k_{b}}+3k_{b})}{k_{a}} \rangle \\ 
    - \langle \frac{a_{k_{a}k_{b}}+b_{k_{a}k_{b}}+k_{b}}{k_{a}} \rangle \frac{a_{k_{a}k_{b}}+\langle k_{b} \rangle }{a_{k_{a}k_{b}}+b_{k_{a}k_{b}}+2\langle k_{b}\rangle} - \langle \frac{k_{b}}{k_{a}} \rangle  \frac{a_{k_{a}k_{b}}+\langle k_{b}\rangle}{a_{k_{a}k_{b}}+b_{k_{a}k_{b}}+\langle k_{b}\rangle}  \Bigg).
\end{split}
\label{app-X}
\end{equation}
Note that all terms in \cref{app-X} are averaged over the joint positive and negative degree distribution $P_{k_{a}k_{b}}$.

We can now apply the Lagrange method to maximise vote-shares $X_{A}$ against a passive controller B. The Lagrangian is derived as $\mathcal{L} = X_{A} + \lambda (\sum_{k_{a}k{b}} P_{k_{a}k_{b}} a_{k_{a}k_{b}} - \langle a_{k_{a}k_{b}} \rangle N) $, where $\lambda$ is the Lagrangian multiplier. Differentiating $\mathcal{L}$ wrt allocations $a_{k_{a}k_{b}}$ we obtain

\begin{equation}
    \begin{split}
    \frac{\partial \mathcal{L}}{\partial a_{k_{a}k_{b}}} = \frac{1}{k_{a}} - \frac{2a_{k_{a}k_{b}}+b_{k_{a}k_{b}}+3k_{b}}{\langle a_{k_{a}k_{b}} \rangle+ \langle b_{k_{a}k_{b}} \rangle +2\langle k_{b} \rangle} \frac{1}{k_{a}} + 2 \frac{\langle a_{k_{a}k_{b} \rangle}+\langle k_{b} \rangle }{(\langle a_{k_{a}k_{b}} \rangle + \langle b_{k_{a}k_{b} \rangle}+2\langle k_{b} \rangle)^2} \frac{a_{k_{a}k_{b}}+b_{k_{a}k_{b}}+2k_{b}}{k_{a}} \\
    - \frac{\langle a_{k_{a}k_{b}} \rangle +2\langle k_{b} \rangle}{\langle a_{k_{a}k_{b}}\rangle + \langle b_{k_{a}k_{b}} \rangle+2\langle k_{b} \rangle} \frac{1}{k_{a}} + \lambda = 0.
    \end{split}
\end{equation}

Solving for $a_{k_{a}k_{b}}$ gives us,

\begin{equation}
    \begin{split}
        \implies a_{k_{a}k_{b}} = \frac{1}{2} \Bigg( \frac{\langle a_{k_{a}k_{b}} \rangle - \langle b_{k_{a}k_{b}}\rangle}{\langle b_{k_{a}k_{b}} \rangle +\langle k_{b} \rangle} b_{k_{a}k_{b}} + \frac{\langle a_{k_{a}k_{b}} \rangle - 3\langle b_{k_{a}k_{b}} \rangle -2 \langle k_{b} \rangle}{\langle b_{k_{a}k_{b}} \rangle +\langle k_{b} \rangle} k_{b} \\ + \langle a_{k_{a}k_{b}}\rangle + \langle b_{k_{a}k_{b}}\rangle+2\langle k_{b} \rangle + \langle k_{a} \rangle \frac{\lambda (\langle a_{k_{a}k_{b}}\rangle + \langle b_{k_{a}k_{b}} \rangle+2\langle k_{b} \rangle)^2}{\langle b_{k_{a}k_{b}\rangle+\langle k_{b} \rangle}} \Bigg),
  \end{split}
  \label{allo}
\end{equation}
which still contains the Lagrangian multiplier $\lambda$. To appropriately deal with $\lambda$ we average over \cref{allo} and assume the budget per node $\langle a_{k_{a}k_{b}} \rangle$ is sufficiently large. Therefore $\frac{\lambda (\langle a_{k_{a}k_{b}}\rangle + \langle b_{k_{a}k_{b}} \rangle+2\langle k_{b} \rangle)^2}{\langle b_{k_{a}k_{b}\rangle+\langle k_{b} \rangle}} \rightarrow 0$, which finally gives us the expression for the optimal allocation,

\begin{equation}
        a_{k_{a}k_{b}}^* = \frac{1}{2} \Bigg( \frac{\langle a_{k_{a}k_{b}}\rangle - \langle b_{k_{a}k_{b}}\rangle}{\langle b_{k_{a}k_{b}} \rangle +\langle k_{b} \rangle}  b_{k_{a}k_{b}} + \frac{\langle a_{k_{a}k_{b}} \rangle - 3\langle b_{k_{a}k_{b}} \rangle -2 \langle k_{b} \rangle}{\langle b_{k_{a}k_{b}} \rangle +\langle k_{b} \rangle} k_{b} + \langle a_{k_{a}k_{b}}\rangle + \langle b_{k_{a}k_{b}}\rangle+2\langle k_{b} \rangle \Bigg).
  \label{allo-final}
\end{equation}

\section{Uniformly distributed negative edges and uniform adversarial allocations}
\label{gain0}

When a controller cannot observe negative edges, the expression for optimal allocation is given as,

\begin{align}
    a_{k} = \frac{1}{2} \Bigg( \frac{\langle a_{k}\rangle - \langle b_{k}\rangle}{\langle b_{k} \rangle}  b_{k} + \langle a_{k}\rangle + \langle b_{k}\rangle \Bigg).
    \label{pos-allo}
\end{align}
where $k = k_{a}+k_{b}$.

The final vote-share in this case $X_{A}^{(+)}$ is obtained by replacing \cref{pos-allo} in \cref{app-X}. 
Gain in vote-shares can therefore be quantified as, 

\begin{align*}
    X_{A} - X_{A}^{(+)} &= \frac{1}{\langle a_{k_{a}k_{b}} \rangle + \langle b_{k_{a}k_{b}} \rangle +2\langle k_{b} \rangle}\Big( (\langle b_{k_{a}k_{b}} \rangle+\langle k_{b} \rangle) \langle \frac{a_{k_{a}k_{b}}-a_{k}}{k_{a}} \rangle \\
    & + \frac{\langle a_{k_{a}k_{b}} \rangle +\langle k_{b} \rangle}{\langle a_{k_{a}k_{b}} \rangle + \langle a_{k_{a}k_{b}} \rangle + 2\langle k_{b} \rangle} \langle \frac{(a_{k_{a}k_{b}}-a_{k})(a_{k_{a}k_{b}}+a_{k}+2\langle b_{k_{a}k_{b}} \rangle+4k_{b})}{k_{a}} \rangle \\
    & - \langle \frac{(a_{k_{a}k_{b}}-a_{k})(a_{k_{a}k_{b}}+a_{k}+ \langle b_{k_{a}k_{b}} \rangle +3k_{b})}{k_{a}} \rangle \Big).
\end{align*}

Furthermore, the term $a_{k_{a}k_{b}} - a_{k}$ in the above expression can be derived using \cref{allo-final,pos-allo} as, 
\begin{align}
    a_{k_{a}k_{b}} - a_{k}  =  \Big( 1 - \frac{\langle a_{k_{a}k_{b}} \rangle - \langle b_{k_{a}k_{b}} \rangle }{2\langle b_{k_{a}k_{b}} \rangle}\frac{b_{k_{a}k{b}}}{\langle b_{k_{a}k_{b}} \rangle+\langle k_{b} \rangle} \Big) \langle k_{b} \rangle + \frac{\langle a_{k_{a}k_{b}} \rangle-3\langle b_{k_{a}k_{b}} \rangle-2\langle k_{b} \rangle}{2(\langle b_{k_{a}k_{b}} \rangle+\langle k_{b} \rangle)} k_{b}.
    \label{diff-allo}
\end{align}

We consider networks with regular negative graphs, $k_{b}=\langle k_{b} \rangle$ where an adversary uniformly targets the network, $b_{k_{a}k_{b}} = \langle b_{k_{a}k_{b}} \rangle$. The above relations further simplify \cref{diff-allo} as,

\begin{align*}
    a_{k_{a}k_{b}} - a_{k} &= \Bigg( \Big( 1 - \frac{\langle a_{k_{a}k_{b}} \rangle- \langle b_{k_{a}k_{b}} \rangle}{2(\langle b_{k_{a}k_{b}} \rangle+\langle k_{b} \rangle)} \Big) + \frac{\langle a_{k_{a}k_{b}} \rangle -3 \langle b_{k_{a}k_{b}} \rangle -2\langle k_{b} \rangle}{2(\langle b_{k_{a}k_{b}} \rangle +\langle k_{b} \rangle)} \Bigg) \langle k_{b} \rangle=0.
\end{align*}

Therefore, it follows that gain $ X_{A} - X_{A}^{(+)}  = 0$, against an adversary targeting all nodes uniformly, in networks with regular negative components.

%% file: main.bbl
\begin{thebibliography}{}

\bibitem[Acemo{\u{g}}lu et~al., 2013]{acemouglu2013opinion}
Acemo{\u{g}}lu, D., Como, G., Fagnani, F., and Ozdaglar, A. (2013).
\newblock Opinion fluctuations and disagreement in social networks.
\newblock {\em Mathematics of Operations Research}, 38(1):1--27.

\bibitem[Bae and Lee, 2012]{bae2012sentiment}
Bae, Y. and Lee, H. (2012).
\newblock Sentiment analysis of twitter audiences: Measuring the positive or negative influence of popular twitterers.
\newblock {\em Journal of the American Society for Information Science and technology}, 63(12):2521--2535.

\bibitem[Banerjee et~al., 2020]{banerjee2020survey}
Banerjee, S., Jenamani, M., and Pratihar, D.~K. (2020).
\newblock A survey on influence maximization in a social network.
\newblock {\em Knowledge and Information Systems}, 62(9):3417--3455.

\bibitem[Barab{\'a}si, 2013]{barabasi2013network}
Barab{\'a}si, A.-L. (2013).
\newblock Network science.
\newblock {\em Philosophical Transactions of the Royal Society A: Mathematical, Physical and Engineering Sciences}, 371(1987):20120375.

\bibitem[Barrat et~al., 2008]{barrat2008dynamical}
Barrat, A., Barthelemy, M., and Vespignani, A. (2008).
\newblock {\em Dynamical processes on complex networks}.
\newblock Cambridge university press.

\bibitem[Bell, 1992]{bell1992note}
Bell, F.~K. (1992).
\newblock A note on the irregularity of graphs.
\newblock {\em Linear Algebra and its Applications}, 161:45--54.

\bibitem[Borgs et~al., 2010]{borgs2010novel}
Borgs, C., Chayes, J., Kalai, A.~T., Malekian, A., and Tennenholtz, M. (2010).
\newblock A novel approach to propagating distrust.
\newblock In {\em International Workshop on Internet and Network Economics}, pages 87--105. Springer.

\bibitem[Braha and de~Aguiar, 2017]{PMID:28542409}
Braha, D. and de~Aguiar, M. A.~M. (2017).
\newblock Voting contagion: Modeling and analysis of a century of u.s. presidential elections.
\newblock {\em PloS one}, 12(5):e0177970.

\bibitem[Castellano et~al., 2009]{castellano2009statistical}
Castellano, C., Fortunato, S., and Loreto, V. (2009).
\newblock Statistical physics of social dynamics.
\newblock {\em Reviews of modern physics}, 81(2):591.

\bibitem[Chakraborty et~al., 2019]{chakraborty2019competitive}
Chakraborty, S., Stein, S., Brede, M., Swami, A., de~Mel, G., and Restocchi, V. (2019).
\newblock Competitive influence maximisation using voting dynamics.
\newblock In {\em Proceedings of the 2019 IEEE/ACM International Conference on Advances in Social Networks Analysis and Mining}, pages 978--985.

\bibitem[Chen and He, 2015]{chen2015influence}
Chen, S. and He, K. (2015).
\newblock Influence maximization on signed social networks with integrated pagerank.
\newblock In {\em 2015 IEEE International Conference on Smart City/SocialCom/SustainCom (SmartCity)}, pages 289--292. IEEE.

\bibitem[Chen et~al., 2011]{chen2011influence}
Chen, W., Collins, A., Cummings, R., Ke, T., Liu, Z., Rincon, D., Sun, X., Wang, Y., Wei, W., and Yuan, Y. (2011).
\newblock Influence maximization in social networks when negative opinions may emerge and propagate.
\newblock In {\em Proceedings of the 2011 siam international conference on data mining}, pages 379--390. SIAM.

\bibitem[Chen et~al., 2018]{chen2018negative}
Chen, Y., Li, H., and Qu, Q. (2018).
\newblock Negative-aware influence maximization on social networks.
\newblock In {\em Proceedings of the AAAI Conference on Artificial Intelligence}, volume~32.

\bibitem[Cheng et~al., 2020]{cheng2020outbreak}
Cheng, C.-H., Kuo, Y.-H., and Zhou, Z. (2020).
\newblock Outbreak minimization vs influence maximization: an optimization framework.
\newblock {\em BMC Medical Informatics and Decision Making}, 20(1):1--13.

\bibitem[Clifford and Sudbury, 1973]{clifford1973model}
Clifford, P. and Sudbury, A. (1973).
\newblock A model for spatial conflict.
\newblock {\em Biometrika}, 60(3):581--588.

\bibitem[Deffuant et~al., 2000]{deffuant2000mixing}
Deffuant, G., Neau, D., Amblard, F., and Weisbuch, G. (2000).
\newblock Mixing beliefs among interacting agents.
\newblock {\em Advances in Complex Systems}, 3(01n04):87--98.

\bibitem[Domingos and Richardson, 2001]{domingos2001mining}
Domingos, P. and Richardson, M. (2001).
\newblock Mining the network value of customers.
\newblock In {\em Proceedings of the seventh ACM SIGKDD international conference on Knowledge discovery and data mining}, pages 57--66.

\bibitem[Easley and Kleinberg, 2010]{easley2010networks}
Easley, D. and Kleinberg, J. (2010).
\newblock {\em Networks, crowds, and markets}, volume~8.
\newblock Cambridge university press Cambridge.

\bibitem[Erd{\H{o}}s and R{\'e}nyi, 1960]{erdHos1960evolution}
Erd{\H{o}}s, P. and R{\'e}nyi, A. (1960).
\newblock On the evolution of random graphs.
\newblock {\em Publ. Math. Inst. Hung. Acad. Sci}, 5(1):17--60.

\bibitem[Galam, 1999]{galam1999application}
Galam, S. (1999).
\newblock Application of statistical physics to politics.
\newblock {\em Physica A: Statistical mechanics and its applications}, 274(1-2):132--139.

\bibitem[Gambaro and Crokidakis, 2017]{gambaro2017influence}
Gambaro, J.~P. and Crokidakis, N. (2017).
\newblock The influence of contrarians in the dynamics of opinion formation.
\newblock {\em Physica A: Statistical Mechanics and its Applications}, 486:465--472.

\bibitem[Girdhar and Bharadwaj, 2016]{girdhar2016signed}
Girdhar, N. and Bharadwaj, K. (2016).
\newblock Signed social networks: a survey.
\newblock In {\em International Conference on Advances in Computing and Data Sciences}, pages 326--335. Springer.

\bibitem[Guha et~al., 2004]{guha2004propagation}
Guha, R., Kumar, R., Raghavan, P., and Tomkins, A. (2004).
\newblock Propagation of trust and distrust.
\newblock In {\em Proceedings of the 13th international conference on World Wide Web}, pages 403--412.

\bibitem[Harrigan and Yap, 2017]{harrigan2017avoidance}
Harrigan, N. and Yap, J. (2017).
\newblock Avoidance in negative ties: Inhibiting closure, reciprocity, and homophily.
\newblock {\em Social Networks}, 48:126--141.

\bibitem[He et~al., 2019]{he2019information}
He, X., Du, H., Feldman, M.~W., and Li, G. (2019).
\newblock Information diffusion in signed networks.
\newblock {\em PloS one}, 14(10):e0224177.

\bibitem[Hegselmann et~al., 2002]{hegselmann2002opinion}
Hegselmann, R., Krause, U., et~al. (2002).
\newblock Opinion dynamics and bounded confidence models, analysis, and simulation.
\newblock {\em Journal of artificial societies and social simulation}, 5(3).

\bibitem[Holley and Liggett, 1975]{holley1975ergodic}
Holley, R.~A. and Liggett, T.~M. (1975).
\newblock Ergodic theorems for weakly interacting infinite systems and the voter model.
\newblock {\em The annals of probability}, pages 643--663.

\bibitem[Hosseini-Pozveh et~al., 2019]{hosseini2019assessing}
Hosseini-Pozveh, M., Zamanifar, K., and Naghsh-Nilchi, A.~R. (2019).
\newblock Assessing information diffusion models for influence maximization in signed social networks.
\newblock {\em Expert Systems with Applications}, 119:476--490.

\bibitem[Hu, 2017]{hucompeting}
Hu, H. (2017).
\newblock Competing opinion diffusion on social networks.
\newblock {\em Royal Society open science 4. 11}, 11.

\bibitem[Jackson, 2010]{jackson2010social}
Jackson, M.~O. (2010).
\newblock {\em Social and economic networks}.
\newblock Princeton university press.

\bibitem[Jendoubi et~al., 2016]{jendoubi2016maximizing}
Jendoubi, S., Martin, A., Li{\'e}tard, L., Hadji, H.~B., and Yaghlane, B.~B. (2016).
\newblock Maximizing positive opinion influence using an evidential approach.
\newblock In {\em Uncertainty Modelling in Knowledge Engineering and Decision Making: Proceedings of the 12th International FLINS Conference}, pages 168--174. World Scientific.

\bibitem[Ju et~al., 2020]{ju2020new}
Ju, W., Chen, L., Li, B., Liu, W., Sheng, J., and Wang, Y. (2020).
\newblock A new algorithm for positive influence maximization in signed networks.
\newblock {\em Information Sciences}, 512:1571--1591.

\bibitem[Kempe et~al., 2003]{kempe2003maximizing}
Kempe, D., Kleinberg, J., and Tardos, {\'E}. (2003).
\newblock Maximizing the spread of influence through a social network.
\newblock In {\em Proceedings of the ninth ACM SIGKDD international conference on Knowledge discovery and data mining}, pages 137--146.

\bibitem[Kiss et~al., 2017]{kiss2017mathematics}
Kiss, I.~Z., Miller, J.~C., Simon, P.~L., et~al. (2017).
\newblock Mathematics of epidemics on networks.
\newblock {\em Cham: Springer}, 598.

\bibitem[Krapivsky and Redner, 2003]{krapivsky2003dynamics}
Krapivsky, P.~L. and Redner, S. (2003).
\newblock Dynamics of majority rule in two-state interacting spin systems.
\newblock {\em Physical Review Letters}, 90(23):238701.

\bibitem[Kuhlman et~al., 2013]{kuhlman2013controlling}
Kuhlman, C.~J., Kumar, V.~A., and Ravi, S. (2013).
\newblock Controlling opinion propagation in online networks.
\newblock {\em Computer Networks}, 57(10):2121--2132.

\bibitem[Kumar et~al., 2018]{kumar2018rev2}
Kumar, S., Hooi, B., Makhija, D., Kumar, M., Faloutsos, C., and Subrahmanian, V. (2018).
\newblock Rev2: Fraudulent user prediction in rating platforms.
\newblock In {\em Proceedings of the Eleventh ACM International Conference on Web Search and Data Mining}, pages 333--341. ACM.

\bibitem[Kumar et~al., 2016]{kumar2016edge}
Kumar, S., Spezzano, F., Subrahmanian, V., and Faloutsos, C. (2016).
\newblock Edge weight prediction in weighted signed networks.
\newblock In {\em Data Mining (ICDM), 2016 IEEE 16th International Conference on}, pages 221--230. IEEE.

\bibitem[Leskovec et~al., 2010]{leskovec2010signed}
Leskovec, J., Huttenlocher, D., and Kleinberg, J. (2010).
\newblock Signed networks in social media.
\newblock In {\em Proceedings of the SIGCHI conference on human factors in computing systems}, pages 1361--1370.

\bibitem[Li et~al., 2017]{li2017positive}
Li, D., Wang, C., Zhang, S., Zhou, G., Chu, D., and Wu, C. (2017).
\newblock Positive influence maximization in signed social networks based on simulated annealing.
\newblock {\em Neurocomputing}, 260:69--78.

\bibitem[Li et~al., 2014]{li2014polarity}
Li, D., Xu, Z.-M., Chakraborty, N., Gupta, A., Sycara, K., and Li, S. (2014).
\newblock Polarity related influence maximization in signed social networks.
\newblock {\em PloS one}, 9(7):e102199.

\bibitem[Li et~al., 2011]{li2011strategy}
Li, Q., Braunstein, L.~A., Havlin, S., and Stanley, H.~E. (2011).
\newblock Strategy of competition between two groups based on an inflexible contrarian opinion model.
\newblock {\em Physical Review E}, 84(6):066101.

\bibitem[Li et~al., 2013]{li2013influence}
Li, Y., Chen, W., Wang, Y., and Zhang, Z.-L. (2013).
\newblock Influence diffusion dynamics and influence maximization in social networks with friend and foe relationships.
\newblock In {\em Proceedings of the sixth ACM international conference on Web search and data mining}, pages 657--666.

\bibitem[Li et~al., 2018]{li2018influence}
Li, Y., Fan, J., Wang, Y., and Tan, K.-L. (2018).
\newblock Influence maximization on social graphs: A survey.
\newblock {\em IEEE Transactions on Knowledge and Data Engineering}, 30(10):1852--1872.

\bibitem[Liang et~al., 2019]{liang2019influence}
Liang, W., Shen, C., Li, X., Nishide, R., Piumarta, I., and Takada, H. (2019).
\newblock Influence maximization in signed social networks with opinion formation.
\newblock {\em IEEE Access}, 7:68837--68852.

\bibitem[Liu et~al., 2019]{liu2019influence}
Liu, W., Chen, X., Jeon, B., Chen, L., and Chen, B. (2019).
\newblock Influence maximization on signed networks under independent cascade model.
\newblock {\em Applied Intelligence}, 49(3):912--928.

\bibitem[Lynn and Lee, 2016]{lynn2016maximizing}
Lynn, C. and Lee, D.~D. (2016).
\newblock Maximizing influence in an ising network: A mean-field optimal solution.
\newblock {\em Advances in neural information processing systems}, 29.

\bibitem[Masuda, 2013]{masuda2013voter}
Masuda, N. (2013).
\newblock Voter models with contrarian agents.
\newblock {\em Physical Review E}, 88(5):052803.

\bibitem[Masuda, 2015]{masuda2015opinion}
Masuda, N. (2015).
\newblock Opinion control in complex networks.
\newblock {\em New Journal of Physics}, 17(3):033031.

\bibitem[Masuda et~al., 2010]{masuda2010heterogeneous}
Masuda, N., Gibert, N., and Redner, S. (2010).
\newblock Heterogeneous voter models.
\newblock {\em Physical Review E}, 82(1):010103.

\bibitem[Mobilia, 2003]{mobilia2003does}
Mobilia, M. (2003).
\newblock Does a single zealot affect an infinite group of voters?
\newblock {\em Physical review letters}, 91(2):028701.

\bibitem[Mobilia, 2015]{mobilia2015nonlinear}
Mobilia, M. (2015).
\newblock Nonlinear q-voter model with inflexible zealots.
\newblock {\em Physical Review E}, 92(1):012803.

\bibitem[Mobilia et~al., 2007]{mobilia2007role}
Mobilia, M., Petersen, A., and Redner, S. (2007).
\newblock On the role of zealotry in the voter model.
\newblock {\em Journal of Statistical Mechanics: Theory and Experiment}, 2007(08):P08029.

\bibitem[Myers et~al., ]{myers1996social}
Myers, D.~G. et~al.
\newblock Social psychology.

\bibitem[Newman, 2018]{newman2018networks}
Newman, M. (2018).
\newblock {\em Networks}.
\newblock Oxford university press.

\bibitem[Newman et~al., 2011]{newman2011structure}
Newman, M., Barab{\'a}si, A.-L., and Watts, D.~J. (2011).
\newblock {\em The structure and dynamics of networks}.
\newblock Princeton university press.

\bibitem[Newman, 2003]{newman2003structure}
Newman, M.~E. (2003).
\newblock The structure and function of complex networks.
\newblock {\em SIAM review}, 45(2):167--256.

\bibitem[Noorazar, 2020]{noorazar2020recent}
Noorazar, H. (2020).
\newblock Recent advances in opinion propagation dynamics: a 2020 survey.
\newblock {\em The European Physical Journal Plus}, 135(6):1--20.

\bibitem[Offer, 2021]{offer2021negative}
Offer, S. (2021).
\newblock Negative social ties: Prevalence and consequences.
\newblock {\em Annual Review of Sociology}, 47.

\bibitem[Pastor-Satorras et~al., 2015]{pastor2015epidemic}
Pastor-Satorras, R., Castellano, C., Van~Mieghem, P., and Vespignani, A. (2015).
\newblock Epidemic processes in complex networks.
\newblock {\em Reviews of modern physics}, 87(3):925.

\bibitem[Perrin, 2015]{perrin2015social}
Perrin, A. (2015).
\newblock Social media usage.
\newblock {\em Pew research center}, 125:52--68.

\bibitem[Pfeffer et~al., 2014]{pfeffer2014understanding}
Pfeffer, J., Zorbach, T., and Carley, K.~M. (2014).
\newblock Understanding online firestorms: Negative word-of-mouth dynamics in social media networks.
\newblock {\em Journal of Marketing Communications}, 20(1-2):117--128.

\bibitem[Prakash et~al., 2012]{prakash2012winner}
Prakash, B.~A., Beutel, A., Rosenfeld, R., and Faloutsos, C. (2012).
\newblock Winner takes all: competing viruses or ideas on fair-play networks.
\newblock In {\em Proceedings of the 21st international conference on World Wide Web}, pages 1037--1046.

\bibitem[Ranganath et~al., 2016]{ranganath2016understanding}
Ranganath, S., Hu, X., Tang, J., and Liu, H. (2016).
\newblock Understanding and identifying advocates for political campaigns on social media.
\newblock In {\em Proceedings of the Ninth ACM International Conference on Web Search and Data Mining}, pages 43--52.

\bibitem[Redner, 2019]{redner2019reality}
Redner, S. (2019).
\newblock Reality-inspired voter models: A mini-review.
\newblock {\em Comptes Rendus Physique}, 20(4):275--292.

\bibitem[Rogers, 2010]{rogers2010diffusion}
Rogers, E.~M. (2010).
\newblock {\em Diffusion of innovations}.
\newblock Simon and Schuster.

\bibitem[Rombach et~al., 2017]{rombach2017core}
Rombach, P., Porter, M.~A., Fowler, J.~H., and Mucha, P.~J. (2017).
\newblock Core-periphery structure in networks (revisited).
\newblock {\em SIAM review}, 59(3):619--646.

\bibitem[Romero~Moreno et~al., 2021]{romero2021shadowing}
Romero~Moreno, G., Chakraborty, S., and Brede, M. (2021).
\newblock Shadowing and shielding: Effective heuristics for continuous influence maximisation in the voting dynamics.
\newblock {\em Plos one}, 16(6):e0252515.

\bibitem[Shen et~al., 2015]{shen2015influence}
Shen, C., Nishide, R., Piumarta, I., Takada, H., and Liang, W. (2015).
\newblock Influence maximization in signed social networks.
\newblock In {\em International Conference on Web Information Systems Engineering}, pages 399--414. Springer.

\bibitem[Sood and Redner, 2005]{sood2005voter}
Sood, V. and Redner, S. (2005).
\newblock Voter model on heterogeneous graphs.
\newblock {\em Physical review letters}, 94(17):178701.

\bibitem[Srivastava et~al., 2015]{srivastava2015social}
Srivastava, A., Chelmis, C., and Prasanna, V.~K. (2015).
\newblock Social influence computation and maximization in signed networks with competing cascades.
\newblock In {\em Proceedings of the 2015 IEEE/ACM International Conference on Advances in Social Networks Analysis and Mining 2015}, pages 41--48.

\bibitem[Tripathy et~al., 2010]{tripathy2010study}
Tripathy, R.~M., Bagchi, A., and Mehta, S. (2010).
\newblock A study of rumor control strategies on social networks.
\newblock In {\em Proceedings of the 19th ACM international conference on Information and knowledge management}, pages 1817--1820.

\bibitem[Watts and Dodds, 2007]{watts2007influentials}
Watts, D.~J. and Dodds, P.~S. (2007).
\newblock Influentials, networks, and public opinion formation.
\newblock {\em Journal of consumer research}, 34(4):441--458.

\bibitem[Watts et~al., 2007]{watts2007viral}
Watts, D.~J., Peretti, J., and Frumin, M. (2007).
\newblock {\em Viral marketing for the real world}.
\newblock Harvard Business School Pub. Boston.

\bibitem[Wilder et~al., 2018]{wilder2018end}
Wilder, B., Onasch-Vera, L., Hudson, J., Luna, J., Wilson, N., Petering, R., Woo, D., Tambe, M., and Rice, E. (2018).
\newblock End-to-end influence maximization in the field.
\newblock In {\em AAMAS}, volume~18, pages 1414--1422.

\bibitem[Yildiz et~al., 2013]{yildiz2013binary}
Yildiz, E., Ozdaglar, A., Acemoglu, D., Saberi, A., and Scaglione, A. (2013).
\newblock Binary opinion dynamics with stubborn agents.
\newblock {\em ACM Transactions on Economics and Computation (TEAC)}, 1(4):1--30.

\bibitem[Zhong et~al., 2005]{zhong2005effects}
Zhong, L.-X., Zheng, D.-F., Zheng, B., and Hui, P. (2005).
\newblock Effects of contrarians in the minority game.
\newblock {\em Physical Review E}, 72(2):026134.

\end{thebibliography}
